\documentclass[10pt, conference, compsocconf]{IEEEtran}
\usepackage{cite}
\usepackage{amsmath,amssymb,amsfonts}
\usepackage{algorithmic}
\usepackage{graphicx}
\usepackage{textcomp}
\usepackage{xcolor}

\usepackage{booktabs}
\usepackage[lined,boxruled]{algorithm2e}
\usepackage{braket}
\usepackage{mathtools}
\usepackage{subcaption}
\usepackage{bbm}
\usepackage{comment}
\usepackage[hyphens]{url}

\def\BibTeX{{\rm B\kern-.05em{\sc i\kern-.025em b}\kern-.08em
    T\kern-.1667em\lower.7ex\hbox{E}\kern-.125emX}}

\begin{document}

\title{Optimized Quantum Program Execution Ordering to Mitigate Errors in\\ Simulations of Quantum Systems}

\author{\IEEEauthorblockN{Teague Tomesh\IEEEauthorrefmark{1}\IEEEauthorrefmark{3}\IEEEauthorrefmark{7},
Kaiwen Gui\IEEEauthorrefmark{2}\IEEEauthorrefmark{7},
Pranav Gokhale\IEEEauthorrefmark{3}, 
Yunong Shi\IEEEauthorrefmark{4},
Frederic T. Chong\IEEEauthorrefmark{5}\IEEEauthorrefmark{3},\\
Margaret Martonosi\IEEEauthorrefmark{1}, and
Martin Suchara\IEEEauthorrefmark{6}\IEEEauthorrefmark{2}}

\IEEEauthorblockA{\IEEEauthorrefmark{1}Department of Computer Science,
Princeton University,
Princeton, USA}
\IEEEauthorblockA{\IEEEauthorrefmark{2}Pritzker School of Molecular Engineering,
University of Chicago,
Chicago, USA}
\IEEEauthorblockA{\IEEEauthorrefmark{3}Super.tech, Chicago, USA}
\IEEEauthorblockA{\IEEEauthorrefmark{4}Amazon Braket, New York, USA}
\IEEEauthorblockA{\IEEEauthorrefmark{5}Department of Computer Science,
University of Chicago,
Chicago, USA}
\IEEEauthorblockA{\IEEEauthorrefmark{6}Mathematics and Computer Science Division,
Argonne National Laboratory,
Lemont, USA}
\IEEEauthorblockA{\IEEEauthorrefmark{7}These authors contributed equally to this work; ttomesh@princeton.edu, kgui@uchicago.edu}
}

\maketitle
\thispagestyle{plain}
\pagestyle{plain}

\begin{abstract}
Simulating the time evolution of a physical system at quantum mechanical levels of detail --- known as Hamiltonian Simulation (HS) --- is an important and interesting problem across physics and chemistry. For this task, algorithms that run on quantum computers are known to be exponentially faster than classical algorithms; in fact, this application motivated Feynman to propose the construction of quantum computers. Nonetheless, there are challenges in reaching this performance potential.   

Prior work has focused on compiling circuits (quantum programs) for HS with the goal of maximizing either accuracy or gate cancellation. Our work proposes a compilation strategy that simultaneously advances both goals. At a high level, we use classical optimizations such as graph coloring and travelling salesperson to order the execution of quantum programs. Specifically, we group together mutually commuting terms in the Hamiltonian (a matrix characterizing the quantum mechanical system) to improve the accuracy of the simulation. We then rearrange the terms within each group to maximize gate cancellation in the final quantum circuit. These optimizations work together to improve HS performance and result in an average 40\% reduction in circuit depth. This work advances the frontier of HS which in turn can advance physical and chemical modeling in both basic and applied sciences.
\end{abstract}

\begin{IEEEkeywords}
quantum computing; compilation; program ordering; Hamiltonian simulation

\end{IEEEkeywords}

\section{Introduction} \label{sec:introduction}
The development of quantum computers is advancing rapidly. During the last decade, quantum computing (QC) systems comprised of tens of qubits were brought online for the first time \cite{arute2019quantum, corcoles2019challenges, wright2019benchmarking}. Although information processing on these devices is noisy and error-prone, successful execution of several quantum algorithms --- including a demonstration of quantum supremacy --- has been achieved \cite{arute2019quantum, barends2016digitized, hempel2018quantum, peruzzo2014variational, tomesh2020coreset}. 

This paper focuses on Hamiltonian simulation (HS) which is used for simulating the time evolution dynamics of a physical system at quantum mechanical detail \cite{lloyd1996universal}. This type of simulation was the original motivation behind Feynman's proposal of QC \cite{feynman1982simulating}.  Lloyd \cite{lloyd1996universal} later proved that quantum computers are indeed efficient simulators of quantum systems, implying an exponential separation between quantum and classical algorithms for this problem. When at-scale, fault-tolerant QCs are available, they are expected to efficiently simulate the dynamics of classically intractable chemical and physical systems \cite{childs2018toward, mueck2015quantum}. HS offers the potential to reveal physical and chemical processes important to materials research, pharmaceuticals, and more \cite{aspuru2005simulated, lidar1999calculating, reiher2017elucidating}.  

 \begin{figure}[t]
    \centering
    \includegraphics[width=1.0\columnwidth]{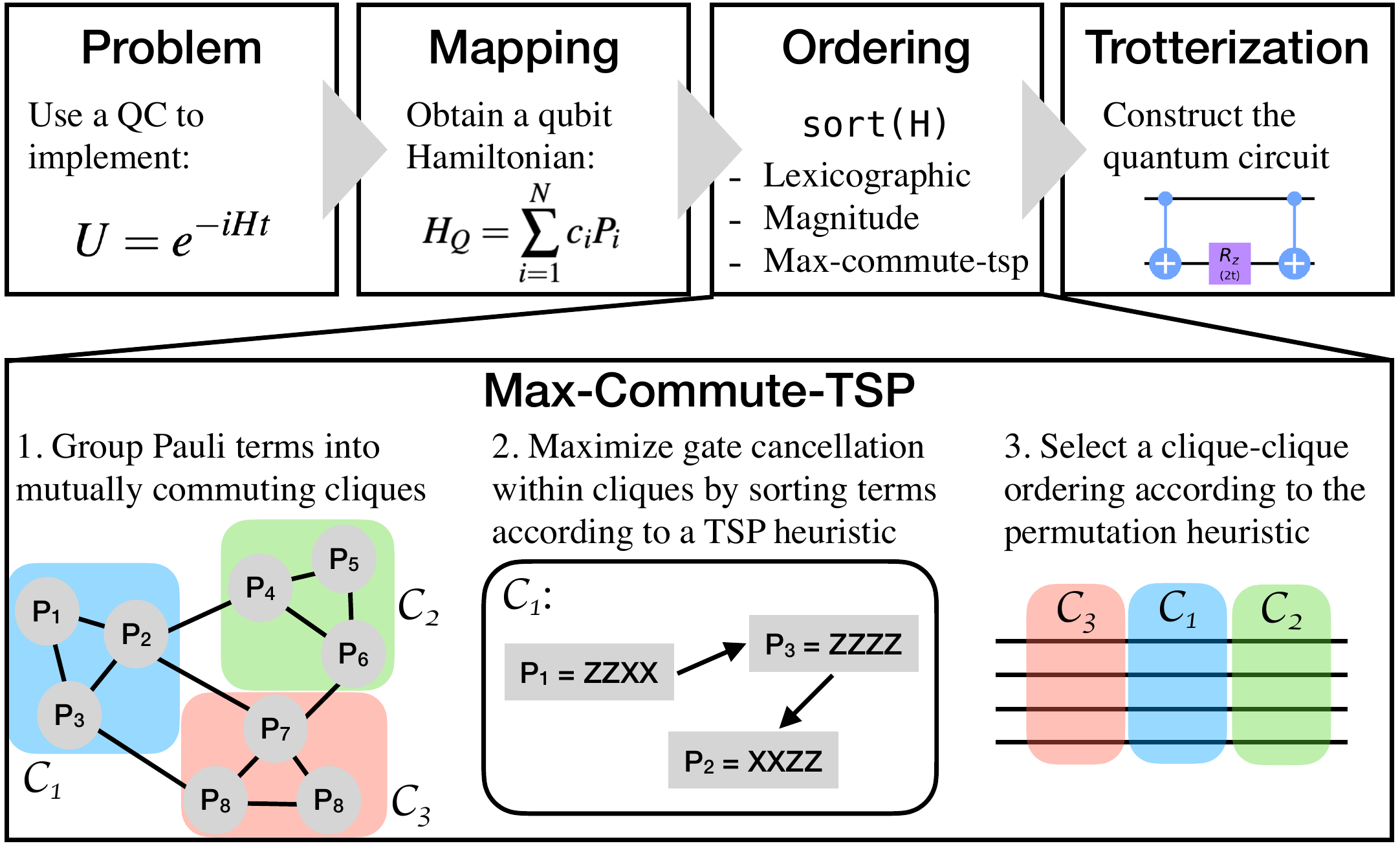}
    \caption{A summary of the Hamiltonian simulation compilation process and the \textit{max-commute-tsp} ordering strategy.}
    \label{fig:big_pic}
\end{figure}

 While HS is promising in theory, key challenges in its uptake lie in demonstrating accurate and tractable HS execution on current QCs.  For current Noisy Intermediate-Scale Quantum (NISQ)~\cite{preskill2018quantum} computers, their limited scale precludes the execution of large HS instances, and their noisy operation impedes accuracy.  
 
 To implement HS on a quantum computer, one needs to specify a particular problem instance (e.g., selecting a specific molecule or defining an optimization objective) in terms of a Hamiltonian---a characteristic matrix describing the system of interest. Then, the quantum circuit (i.e., quantum program) that simulates the system is compiled via three steps: mapping, ordering, and Trotterization, shown in Fig.~\ref{fig:big_pic}. The goal of the {\em mapping} step is to produce a Hamiltonian specified as a sum of Pauli terms (tensor products of Pauli matrices, more details in Sec.~\ref{subsec:dqs_circuits}) which act on the qubits of the quantum computer, although the specifics may vary depending on the use case.
 
 The quantum computer executes the HS by sequentially simulating each Pauli term in the order that it appears in the summation produced during the \textit{mapping} step. While any ordering is theoretically viable, some orderings have very poor performance or accuracy given current QC constraints. Thus, the purpose of the {\em ordering} step is to sort the Pauli terms with the goal of (i) minimizing the depth (i.e., quantum operation count or runtime) of the resulting quantum circuit and (ii) maximizing the accuracy of the simulation. 
 This is analogous to the ordering of floating point operations or basic block scheduling performed by classical compilers~\cite{appel2004modern}.
 Prior work has explored the impact that term order has on the quality of the simulation, and finding tighter bounds on the simulation error remains an open problem~\cite{babbush2015chemical, childs2019faster, tran2020destructive, tranter2019ordering}. Additionally, the impact of term ordering on the gate count requirements for the simulation circuits has also been studied \cite{tranter2018comparison, hastings2014improving}. 
 
 Finally, the last step in the circuit compilation is \textit{Trotterization} which iteratively constructs the quantum circuit according to the Suzuki-Trotter decomposition \cite{suzuki2005}. 
 
 While prior works offered theoretical advances, their term ordering strategies focused solely on either accuracy or gate cancellation. Our work highlights the importance of \textit{both} of these factors and applies classical optimization techniques to compile quantum circuits which are both short and accurate. 
 
 We improve the execution of HS circuits by developing methods to \textbf{mitigate both physical and algorithmic errors}. \textit{Physical} errors stem from the fact that gate operations on NISQ processors are noisy. Two instances of the same HS problem compiled with different term orderings can result in quantum circuits of different lengths. The shorter circuit contains fewer noisy operations and will, therefore, have a higher probability of successful execution. Even in the fault-tolerant (i.e., error-corrected) regime, where gate operations are noiseless, compilations which produce shorter circuits are still desirable since depth is proportional to runtime. \textit{Algorithmic} errors appear in the Trotterization step (covered in detail in Sec.~\ref{subsec:dqs_circuits}) because the continuous time evolution that we wish to simulate must be discretized before it can be implemented on a quantum computer. This discretization is only an approximation of the true evolution. Prior work as well as our results indicate that the order in which the Pauli terms are simulated can significantly impact the error of this approximation.

Our strategy (summarized in Fig.~\ref{fig:big_pic}) for optimizing the program order of the simulation is based on two key insights. First, the algorithmic error associated with the Trotterization step is \textbf{due to non-commuting Pauli terms} within the Hamiltonian. Noticing this, we mitigate the algorithmic error by grouping together Pauli terms which commute with one another. We decide the grouping by constructing a graph that indicates which Pauli terms commute with one another, and then find a minimum clique cover on this graph. The second insight follows from the first, namely, once the Pauli terms are partitioned into mutually commuting groups, the \textbf{terms within each group can be rearranged without incurring any additional algorithmic error}. We choose to arrange the Pauli terms within each group to maximize the amount of gate cancellation in the final circuit. This is accomplished by finding a travelling salesperson path through each group which places similar Pauli terms next to each other, thus increasing the amount of gate cancellation.

These compilation techniques incorporate both classical optimizations and full-stack knowledge from application to hardware which improves the performance of Hamiltonian simulation. This approach has proven successful in previous work targeting the Variational Quantum Eigensolver (a quantum algorithm which can be considered a specialized case of HS~\cite{mcclean2016theory, van2020circuit})~\cite{gokhale2019minimizing, verteletskyi2020measurement, yen2020measuring}. Furthermore, considering the physics at the hardware level --- the QC must discretize the simulation to handle non-commuting Pauli terms --- guides our optimizations at the compiler level: grouping together commuting Pauli terms. Going in the other direction, we use the compiler to optimize for short circuits which fit within the limits set by the capabilities of the underlying hardware.

In this paper we consider simulation and experiments for molecular Hamiltonians because they are a relevant and important application and also easily obtainable via the NIST Chemistry WebBook~\cite{nist1997chemistry} and OpenFermion software package~\cite{mcclean2017openfermion}.

Our contributions in this work include:
\begin{itemize}
    \item A new term ordering strategy, \textit{max-commute-tsp}, which \textbf{simultaneously mitigates both physical and algorithmic errors}.
    \item Simulation and experimental results which demonstrate an average \textbf{40\% reduction in circuit depth} and highlight the \textbf{importance of both gate cancellation and simulation accuracy} for good overall performance.
    \item A general, open-source implementation of HS which can be of use to the QC community as a challenging and practical benchmark.
\end{itemize}

The rest of the paper is organized as follows. An overview of HS is given in Sec.~\ref{sec:dqs_algorithm}. Prior ordering strategies and compilation techniques are discussed in Sec.~\ref{sec:prior_work} and a detailed description of \textit{max-commute-tsp} is presented in Sec.~\ref{sec:max-commute-tsp}. Our benchmarking methodology is given in Sec.~\ref{sec:methodology} and Sec.~\ref{sec:benchmarks} contains the results of our simulations and the evaluations on trapped ion quantum computers. In Sec.~\ref{sec:conclusion} we discuss our results as well as future work on analyzing HS performance in regimes beyond quantum chemistry.

\begin{figure}
    \centering
    \includegraphics[width=\columnwidth]{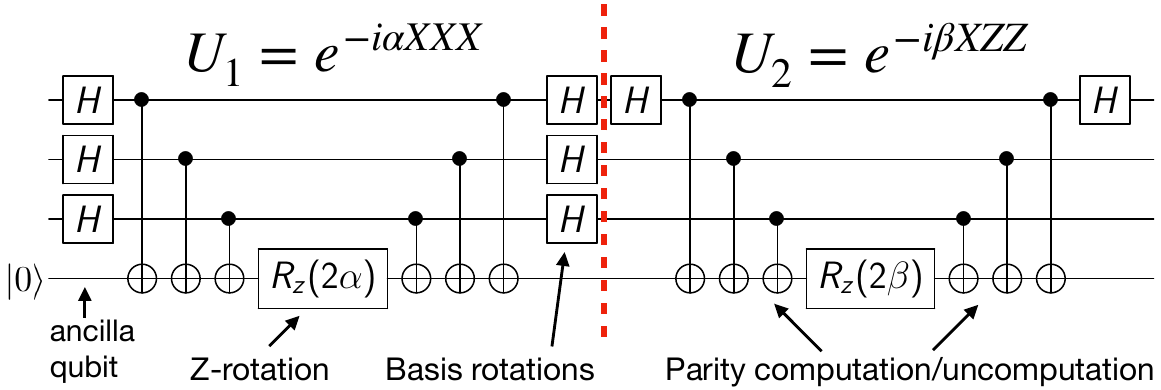}
    \caption{Dissecting a quantum circuit for simulating the evolution of $H = \alpha XXX + \beta XZZ$. The red dotted line separates the two subcircuits implementing each individual term. The full circuit implements the evolution unitary $U = U_2 U_1 = e^{-i\beta XZZ}e^{-i\alpha XXX}$.}
    \label{fig:example_circuit}
\end{figure}
\section{Hamiltonian Simulation}\label{sec:dqs_algorithm}

Simulating quantum systems is a general problem with important applications in physics, chemistry, and biology. Many supercomputer hours are spent simulating different molecules and materials each year~\cite{incite}. With the recent development of quantum computers, and because they are efficient simulators of quantum systems, it is expected that these simulation problems will be among the most promising applications for quantum advantage~\cite{mueck2015quantum}. However, it is likely that the simulation of classically-intractable systems will require a fault-tolerant quantum computer~\cite{childs2018toward}. Despite this, Hamiltonian simulation is still used throughout many near-term NISQ algorithms for machine learning and other optimization applications. For the remainder of this paper, we choose to consider Hamiltonian simulation within the specific context of simulating molecular dynamics. Our compilation methods can be further applied wherever one wishes to simulate a Hamiltonian written as a sum of Pauli terms.

\subsection{The Simulation Problem}
To illustrate Hamiltonian simulation, assume we wish to study a time-dependent quantum system, $\ket{\psi(t)}$, such as an atom or molecule. The time evolution of this system will be described by a matrix, $H$, called the Hamiltonian whose eigenvalues are the allowed energy levels of the system. If the initial state of the system is $\ket{\psi(0)}$, then the state at a later time is given by the equation
\begin{equation} \label{eqn:evolution}
    \ket{\psi(t)} = U\ket{\psi(0)} = e^{-i H t}\ket{\psi(0)}.
\end{equation}
The simulation problem is solved by computing the unitary evolution matrix $U$.
In general, this is a difficult task because the size of the Hamiltonian grows exponentially with the size of the system.

\subsection{Quantum Circuits for Hamiltonian Simulation}\label{subsec:dqs_circuits}
Exponential scaling with system size is characteristic of quantum mechanics, and it is one of the main reasons why classical methods for solving quantum problems are intractable. However, a quantum computer is capable of representing exponentially large state vectors and evolution matrices using polynomially many qubits and gate operations. Below, we describe the compilation process for constructing the simulation circuits.

\textit{\textbf{Mapping --}}
The molecular Hamiltonians considered here are typically written in terms of fermionic operators which must be mapped to operators that act on qubits~\cite{jordan1928pauli, bravyi2002fermionic}. The Hamiltonian can then be written as a sum of Pauli terms:
\begin{equation} \label{eqn:qubitH}
    H = \sum_{i=1}^{N}c_i P_i.
\end{equation}
Here, $c_i \in \mathbb{R}$ and the Pauli terms, $P_i$, are tensor products of the Pauli matrices with length equal to the number of qubits.
\begin{equation}
    P_i = \bigotimes_{n=1}^{n_q}m_n \;\text{  where  }\; m_n \in \{I,X,Y,Z\} \nonumber
\end{equation}

\textit{\textbf{Ordering --}}
A quantum computer simulates the evolution Eq.~\eqref{eqn:evolution} under a Hamiltonian Eq.~\eqref{eqn:qubitH} by sequentially simulating each of the individual terms $c_iP_i$. Any ordering of the Pauli terms is valid, but some orderings are able to mitigate physical and algorithmic errors more effectively than others. Although the simulation error's dependence on term ordering has been studied extensively~\cite{tranter2019ordering, childs2019faster, childs2019theory}, finding tight bounds on the error remains an open problem \cite{poulin2014trotter, babbush2015chemical}. In addition, the gate requirements for Hamiltonian simulation circuits and numerous circuit optimizations have also been studied, and have shown that different orderings can have different amounts of gate cancellation~\cite{childs2018toward, hastings2014improving, raeisi2012quantum}.

In this work, we consider a number of different compilation methods including \textit{lexicographic} \cite{hastings2014improving, tranter2018comparison}, \textit{magnitude} \cite{hastings2014improving, tranter2018comparison}, \textit{depleteGroups}~\cite{tranter2019ordering}, \textit{random}, and our newly-proposed \textit{max-commute-tsp} term orderings. An overview of each of these strategies is given in Sec.~\ref{sec:prior_work} and \textit{max-commute-tsp} is discussed in detail in Sec.~\ref{sec:max-commute-tsp}.

\textit{\textbf{Trotterization --}}
Once the Hamiltonian has been generated and an ordering is selected, a quantum circuit is constructed which implements the evolution unitary $U$ in Eq.~\eqref{eqn:evolution}. Importantly, if all of the terms in $H$ commute with one another, then $U$ can immediately be written as a product of individual terms
\begin{equation}
    \begin{aligned}
        U = &\; e^{-i H t} = e^{-i c_1 P_1 t} e^{-i c_2 P_2 t} \dots e^{-i c_N P_N t} \\ 
            & \;\text{ if}\; P_j P_k = P_k P_j,\; \forall j,k \in [N].
    \end{aligned}\nonumber
\end{equation}
In general, however, real-world Hamiltonians may contain non-commuting terms. In this case, we can use the Suzuki-Trotter decomposition \cite{suzuki2005} to break the evolution of non-commuting terms into many small time steps to approximate the total evolution unitary \cite{whitfield2011simulation}:
\begin{equation} \label{eqn:trotter}
    e^{-i H t} \approx (e^{-i c_1 P_1 \Delta t} e^{-i c_2 P_2 \Delta t} \dots e^{-i c_N P_N \Delta t})^{t/\Delta t} + \textit{O}(t \Delta t).
\end{equation}
We denote $t / \Delta t = r$ as the Trotter number. The product of exponentials in Eq.~\eqref{eqn:trotter} can be represented as a single quantum circuit, repeated $r$ times, while the remaining $O(t \Delta t) = O(t^2 / r)$ terms, referred to as the Trotter error, is the algorithmic error associated with the Trotterization process. As $r \rightarrow \infty$ the Trotter error vanishes and the quantum simulation becomes exact, but this comes at a cost of increasing circuit depth.

Eq.~\eqref{eqn:trotter} is an example of a first-order Trotter decomposition; higher order decompositions also exist which further improve the accuracy of the approximation (see Sec. 4.7 of \cite{nielsen2002quantum}), however, for simplicity we consider only the first-order decomposition in this paper. 

Fig.~\ref{fig:example_circuit} shows a simulation circuit for an example Hamiltonian. Both terms require single-qubit gates to rotate each qubit into the computational basis. The parity of the qubits is then computed by performing a CNOT, controlled by a data qubit and targeting an ancilla qubit, for each Pauli matrix in the Pauli term. Another method for computing the parity uses a ladder of CNOTs between nearest neighbors to compute the parity. We choose to use the former method even though it introduces a single ancilla overhead because it allows for additional gate cancellations which are not possible with the ladder implementation.

\section{Prior Term Ordering Strategies}\label{sec:prior_work}
Prior work has studied the use of commutativity at the circuit level to optimize the gate counts for the Quantum Approximate Optimization Algorithm (QAOA)~\cite{alam2020circuit} and more general quantum circuits~\cite{itoko2019quantum}. The optimizations in \cite{alam2020circuit} are complementary to the techniques discussed in this paper which operate at the abstraction level of Pauli terms to minimize errors before the application is compiled down into a quantum circuit. In addition, we list here a variety of ordering strategies which have been introduced and their impact on Trotter error or gate costs studied in previous work.

\textbf{Lexicographic --}
Prior work has proposed and improved the \textit{lexicographic} ordering which orders Pauli terms alphabetically to achieve high levels of gate cancellation \cite{hastings2014improving, tranter2018comparison}. The \textit{lexicographic} ordering can produce circuits with shorter depth since consecutive Pauli terms which act on the same qubit with the same Pauli matrix result in single- and two-qubit gates cancelling at the interface between the two terms (see Fig.~\ref{fig:example_circuit}). Although this ordering produces short circuits, it does nothing to mitigate Trotter (algorithmic) errors and therefore a larger Trotter number (i.e., longer circuits) will be required to attain high accuracy. Additionally, Sec.~\ref{sec:tsp} introduces a pathological example where the circuits produced by a \textit{lexicographic} ordering are asymptotically equivalent to no gate cancellation at all.

\textbf{Magnitude --}
Prior work also considered sorting the Pauli terms according to the magnitude of their coefficients in Eq.~\eqref{eqn:qubitH} in descending order \cite{hastings2014improving, tranter2018comparison}. Interestingly, the \textit{magnitude} ordering can produce simulation circuits with very low Trotter error, often outperforming the analytically computed bounds on the Trotter error. Tranter et al. \cite{tranter2019ordering} suggest that the superior accuracy of the \textit{magnitude} strategy may be attributable to simulating the terms with large coefficients earlier in the circuit and so they cannot compound any errors that occur later on. Despite the low algorithmic errors of a \textit{magnitude} ordering, the resulting circuits will be quite deep because this ordering does not utilize any information related to gate cancellations between terms.

\textbf{DepleteGroups --}
The \textit{depleteGroups} strategy was proposed by Tranter et al. \cite{tranter2019ordering}, which also partitions the Pauli terms into groups where every term commutes with every other term within the group. 
Once the Pauli terms are grouped into mutually commuting cliques, the final ordering is produced by iteratively selecting the highest magnitude term from each clique until all the groups have been exhausted, which is the opposite of the approach described in Sec.~\ref{sec:max-commute-tsp}.

\textbf{Random --}
We also consider a \textit{random} ordering of the Pauli terms to serve as a baseline for comparison. Random term orderings were also used by Childs et. al.~\cite{childs2019faster} to prove stronger bounds on the size of the Trotter error. 
\section{Max-Commute-TSP} \label{sec:max-commute-tsp}
As shown in Fig.~\ref{fig:big_pic}, the \textit{max-commute-tsp} ordering is composed of three parts. This algorithm sorts the Hamiltonian by first grouping the Pauli terms into mutually commuting cliques. It then orders the terms within each clique according to a travelling salesperson (TSP) heuristic before heuristically selecting a permutation of the cliques. The following sections describe each of these steps and also motivate why \textit{max-commute-tsp} is able to mitigate both the physical and algorithmic errors in Hamiltonian simulation.

\subsection{Term Grouping to Mitigate Trotter Errors}\label{sec:mincliquecover}

The first step in the \textit{max-commute-tsp} ordering is to construct a commutation graph. The nodes of the commutation graph represent the Pauli terms of the Hamiltonian, and an edge exists between every pair of Pauli terms which commute with one another. The terms are then grouped into a minimum number of mutually commuting cliques that cover the entire graph. 

Recall that a quantum computer is only able to carry out the Hamiltonian simulation by Trotterizing the continuous evolution~\cite{suzuki2005}. This Trotterization step is required to deal with non-commuting terms in the Hamiltonian and incurs Trotter errors, as discussed in Sec.~\ref{sec:dqs_algorithm}. In practice, some of the Trotter errors may be avoided by grouping as many commuting terms together as possible. 

\textit{\textbf{Theoretical Analysis of Commutation Groupings --}}
We illustrate the effectiveness of the group commutation ordering strategy with an example Hamiltonian containing two commuting cliques. The analysis can be generalized to more commutation groups by the reader. Consider a Hamiltonian $H= \sum_{i=1}^{k}\alpha_i H_i$, where $a_i$ are real numbers and $H_i$ are simple Hamiltonians that can be mapped to quantum circuits (or diagonalized) directly. Suppose $H$ can be divided into two commuting groups (cliques) $H^c_1 = \sum_{m=1}^{p} H_m$ and $H^c_2 = \sum_{m=p+1}^{k}H_m$, $i.e.$, $[H_m, H_n] = 0$ if $0<m, n \leq p$ or $p< m, n \leq k$. We compare the Trotter error of a term grouping strategy against other orderings below.

For a \textit{group commutation ordering} of $H$, it has been shown that the approximation error (in the additive form) of the Lie-Trotter formula is given by the variation-of-parameters formula \cite{knapp2005basic,dollard1979product, childs2019theory},

\begin{align}
    \delta_{gc} &= e^{- i t(H_1^c)}e^{- i t(H_2^c)}-e^{-i t H} \nonumber\\
    &=\int_0^t d\tau e^{-i(t-\tau)H}[e^{-i\tau H^c_1}, H^c_2]e^{\tau H^c_2}.
    \label{eq:error}
\end{align}

Note that the error form also applies to general $H^g$ where  $H^g=H^g_1+H^g_2$. For simplicity, we denote the integral in Eq.~\eqref{eq:error} as $I(H^g_1, H^g_2$) for general $H^g_1, H^g_2$. Thus, we can simply write the approximation error as $\delta_{gc} = I(H^c_{1}, H^c_2)$.

We are interested primarily in the operator norm ($i.e.$, spectral norm) of $||\delta_{gc}||= ||I(H^c_{1}, H^c_2)||$, which gives the worst-case analysis of the error.\\

We can recursively apply the error formula in Eq.~\eqref{eq:error} to the Lie-Trotter formula for an \textit{arbitrary ordering} of $H$.  Let $\pi$ be a permutation of the set $\{1,..,p\}$ that defines the ordering. First, we can approximate $e^{-i t H}$ by separating $H_{\pi(1)}$ from other terms:
\begin{align*}
    \delta_1 & = e^{-itH_{\pi(1)}}e^{-it\sum_{m=2}^kH_{\pi(m)}}- e^{-i t H}\\
    &=I(H_{\pi(1)}, \sum_{m=2}^kH_{\pi(m)}).
\end{align*}

Then we can recursively repeat the  process for the rest of the Hamiltonian and arrive at the following expression for the approximation error $\delta$ of the Lie-Trotter formula.

\begin{figure}[t]
    \centering
    \includegraphics[width=1.0\columnwidth]{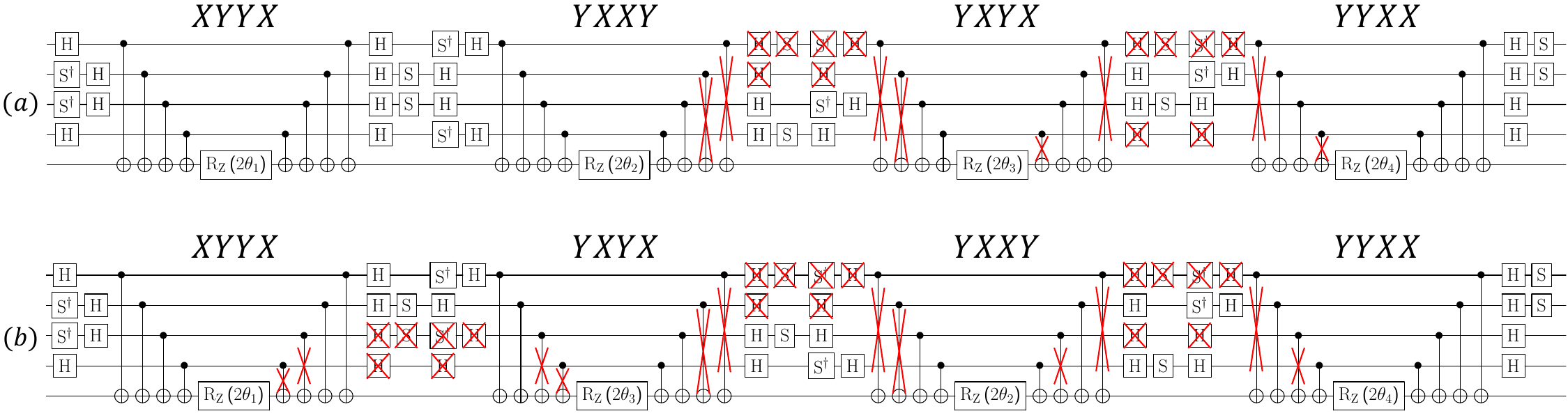}
    \caption{An example showing the gate cancellations that are available with (a) \textit{lexicographic} and (b) TSP orderings.}
    \label{fig:cancellation}
\end{figure}

\begin{align*}
    \delta_{ngc} &= e^{-it H_{\pi(1)}}e^{-it H_{\pi(2)}}...e^{-it H_{\pi(2)}}-e^{-itH}\\
    &= I(H_{\pi(1)}, \sum_{m=2}^pH_{\pi(m)}) + e^{-it H_{\pi(1)}} I(H_{\pi(2)}, \sum_{m=3}^pH_{\pi(m)})\\
    & + e^{-it H_{\pi(1)}}e^{-it H_{\pi(2)}} I(H_{\pi(3)}, \sum_{m=4}^pH_{\pi(m)})+... \\
    &+ e^{-it H_{\pi(1)}}...e^{-it H_{\pi(j)}} I(H_{\pi(j+1)}, \sum_{m=j+2}^pH_{\pi(m)})+...
\end{align*}
Using the triangle inequality and the submultiplicativity of the operator norm, together with the fact that the operator norm of a unitary is 1, we have
\begin{align*}
    ||\delta_{ngc}|| &\approx 
   ||I(H_{\pi(1)}, \sum_{m=2}^pH_{\pi(m)}) +  I(H_{\pi(2)}, \sum_{m=3}^pH_{\pi(m)})\\
    & +  ... + I(H_{\pi(j+1)}, \sum_{m=j+2}^pH_{\pi(m)})+...|| .
\end{align*}

Also we know that $H_{\pi(k)}$ is either in $H^c_1$ or $H^c_2$. Thus, we have
\begin{align*}
    ||\delta_{ngc}||& \approx ||\sum_{\pi(j) \in [1,p]} I(H_{\pi(j)}, H^c_2)||+\\
    &||\sum_{\pi(j) \in [p+1,k]} I(H_{\pi(j)}, H^c_1)||.
\end{align*}

Although we have no proof that $||\delta_{gc}|| < ||\delta_{ngc} ||$ (because we do not have  information about the full commutation relation and magnitude information in $H$), we can, however, make several observations why group commutation ordering is advantageous. First, $||\delta_{gc}||$ in general has a much lower upper bound than does the first term in $||\delta_{ngc}||$. In fact, the upper bound of $||\delta_{gc}||$ does not scale with the number of terms $p$ while $||I(H_{\pi(1)}, \sum_{m=2}^pH_{\pi(m)}) ||$ is of $O(p)$. Second, $||\delta_{gc}||$ does not include the second term in $||\delta_{ngc}||$. Thus, there is strong evidence that \textit{group commutation} ordering has an advantage over naive ordering in terms of eliminating Trotter errors.

\textit{\textbf{Algorithms for Min-Clique-Cover --}}
Partitioning the commutation graph of a quantum Hamiltonian has been proposed for other quantum computing applications, including Hamiltonian simulation~\cite{van2020circuit} and minimizing the number of measurements required in variational algorithms such as the Variational Quantum Eigensolver \cite{gokhale2019minimizing, yen2020measuring, verteletskyi2020measurement}. 

The minimum clique cover problem is NP-Complete \cite{karp1972reducibility}. Fortunately, we do not require the exact min-clique-cover solution for the purpose of this work since approximate solutions are able to effectively mitigate both algorithmic and physical errors. The commutation graphs produced by real-world applications (especially molecular Hamiltonians) tend to be highly structured allowing for reasonably good solutions to be found in time scaling like $O(N^2)$ or $O(N^3)$ for a graph with $N$ nodes~\cite{boppana1992approximating, leighton1979graph, planat2007pauli}.

\subsection{Travelling Salesperson For Gate Cancellation}\label{sec:tsp}
Because the terms within each clique commute with one another, they can be rearranged at will without incurring any additional Trotter error. Therefore, the Pauli terms within cliques can be sorted into an order which maximizes gate cancellation without the worry of degrading the accuracy of the simulation.

We frame the question of ordering the Pauli terms within cliques as an instance of the Travelling Salesperson problem. Recall the example simulation circuit in Fig.~\ref{fig:example_circuit}, we characterize the potential for gate cancellation between two terms using the following principles:
\begin{itemize}
    \item Two CNOT gates with the same target and control qubits can be cancelled so long as no single-qubit gates lie between them. 
    \item Sequential Pauli terms which act on the same qubits with the same Pauli matrix will be able to cancel the basis rotation single-qubit gates between them.
\end{itemize}

According to these gate cancellation principles, it will be favorable to place two Pauli terms next to each other if they operate on the same qubits with the same Pauli matrices. This can be framed as an instance of the Travelling Salesperson problem, where instead of distance travelled the cost function is the number of two-qubit gates in the final circuit.

We focus primarily on the cancellation of two-qubit gates, because they are the dominant source of errors and latency in the prevalent quantum platforms. Typically, CNOT gates have at least 10x lower fidelity and 2--5x longer duration than single-qubit gates \cite{debnath2016demonstration, linke2017experimental}.

\textit{\textbf{Defining the TSP Instance --}}
The TSP objective is to order a set of $k$ Pauli terms, $\{P_1, P_2, ..., P_k\}$, such that the number of CNOT gates in the resulting circuit is minimized.

Every Pauli term within the Hamiltonian corresponds to a subcircuit implemented on the quantum computer. Within a Pauli term, each non-$I$ Pauli matrix generates two identical CNOTs (one for parity-compute, the other for parity-uncompute). For example, the Pauli term: $XZZ$ in Fig.~\ref{fig:example_circuit} has a total of 6 CNOTs in its subcircuit because it has three non-$I$ Pauli matrices. These CNOTs are controlled on the $m$-th qubit (where $m =$ the index of the Pauli matrix within the Pauli term) and target the ancilla qubit. Without any gate cancellation, the number of CNOT gates required to implement the simulation circuit for a Hamiltonian, $H = \sum_{j=1}^{k}c_j P_j$, is
\begin{equation} 2 \sum_{j=1}^k |P_j|_\text{Ham} = 2 \sum_{j=1}^k \sum_{i=1}^N \mathbbm{1}_{P_j[i]\neq I}, \label{eq:cnot_upper_bound} \end{equation}
where $N$ denotes the width of the Pauli terms and Ham refers to Hamming weight.

A good permutation, however, can substantially reduce the number of required CNOTs compared to Eq.~\eqref{eq:cnot_upper_bound}'s upper bound because the CNOT gates between neighboring Pauli term subcircuits can cancel with one another. Using the gate cancellation principles listed above, we define the CNOT distance between two Pauli terms as
\begin{equation}
\begin{split}
    |P_1 - P_2|_\text{CNOT} &:= |P_1 - P_2|_\text{Ham} + \sum_{i \in [N]} \mathbbm{1}_{I \ \neq P_1[i] \neq P_2[i] \neq I} \\
    &= \sum_{i \in [N]} \mathbbm{1}_{P_1[i] \neq P_2[i]}(1 + \mathbbm{1}_{I \not\in \{P_1[i], P_2[i]\}}).
    \label{eq:cnot_distance}
\end{split}
\end{equation}
This distance reports the number of CNOTs needed to implement the transition from the parity-uncompute zone of the $P_1$ subcircuit through the parity-compute zone of the $P_2$ subcircuit, after all possible gates have been cancelled.

The TSP instance is then defined as: given a graph, with nodes representing Pauli terms and edges between nodes weighted according to the CNOT distance, find the shortest cycle which visits each vertex once. To be precise, we actually desire the shortest Hamiltonian path, e.g., we want to visit each Pauli term once, without returning to the start. This is accomplished by generating the TSP cycle and then deleting the most expensive edge in the path.

\textit{\textbf{Approximating TSP --}}
Solving TSP in the most general setting is NP-hard. Moreover, no polynomial-time algorithms exist which are guaranteed to approximate it to a constant ratio \cite{sahni1976p}. In the case of \textit{metric graphs}, however, TSP can be efficiently 1.5-approximated via Christofides' algorithm, meaning that the approximation will return an ordering that requires at most 1.5x as many CNOTs as the optimal ordering~\cite{christofides1976worst}. In addition, Christofides' algorithm is fast, running in $O(k^3)$ time, and it is known to perform well in practice, often attaining near-optimal solutions~\cite{genova2017experimental}.

We now prove that the graph defined by the $|P_1 - P_2|_\text{CNOT}$ distance function is a metric graph. We have already seen that it is symmetric in its arguments (i.e., the graph is undirected), so we need only show that it satisfies the triangle inequality:
\begin{equation}
|P_1 - P_2|_\text{CNOT} + |P_2 - P_3|_\text{CNOT} - |P_1 - P_3|_\text{CNOT} \geq 0 \nonumber .
\end{equation} 

Expanding this expression we obtain:
\begin{equation}
\begin{split}
|P_1 - P_2|_\text{CNOT} + |P_2 - P_3|_\text{CNOT} - |P_1 - P_3|_\text{CNOT}\\
= \sum_{i \in [N]}
\mathbbm{1}_{P_1[i] \neq P_2[i]}(1 + \mathbbm{1}_{I \not\in \{P_1[i], P_2[i]\}}) \\
+ \mathbbm{1}_{P_2[i] \neq P_3[i]}(1 + \mathbbm{1}_{I \not\in \{P_2[i], P_3[i]\}}) \\
- \mathbbm{1}_{P_1[i] \neq P_3[i]}(1 + \mathbbm{1}_{I \not\in \{P_1[i], P_3[i]\}}).
\end{split} \label{eq:big_sum}
\end{equation}
We will prove that each three-term expression in the sum is non-negative for each $i$, so that the full sum must also be non-negative. Note that the third term evaluates to 0, -1, or -2.
\begin{itemize}
    \item If it is 0, the three-term expression is already non-negative since the first two terms are non-negative.
    \item If it is -1, then $P_1[i] \neq P_3[i]$, and one of the two is $I$. We must also have $P_1[i] \neq P_2[i]$ or $P_2[i] \neq P_3[i]$, so the first two terms must sum to at least 1. Thus, the three-term expression is non-negative.
    \item If it is -2, then $P_1[i] \neq P_3[i]$, and neither is $I$. Suppose that $P_1[i] = P_2[i]$; then the first term is +2, and the three-term expression is non-negative. Similarly, if $P_2[i] = P_3[i]$, then the second term is +2, and the three-term expression is non-negative. And if $P_1[i] \neq P_2[i] \neq P_3[i]$, then the first two terms are both at least +1, so the three-term expression is non-negative.
\end{itemize}
Thus, we conclude that Eq.~\eqref{eq:big_sum} is a sum over non-negative numbers which proves that the triangle inequality holds for the CNOT distance. Therefore, our graph is metric, and Christofides' algorithm can be used to efficiently attain a 1.5-approximation to the optimal TSP.

\begin{figure}[h]
  \centering
  \begin{subfigure}[b]{0.135\textwidth}
    \includegraphics[width=\textwidth]{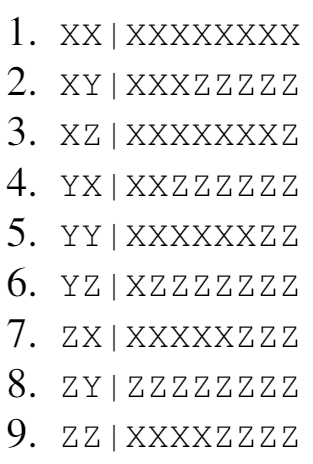}
    \caption{\textit{Lexicographic}}
    \label{fig:lex_nine}
  \end{subfigure}
  \hspace{0.05\textwidth}
  \begin{subfigure}[b]{0.14\textwidth}
    \includegraphics[width=\textwidth]{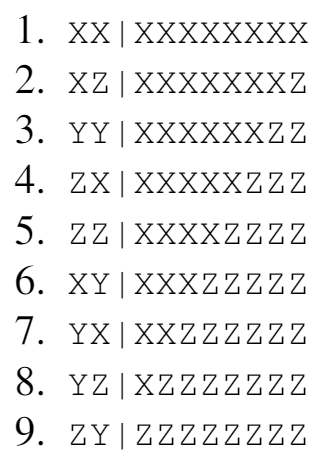}
    \caption{TSP}
    \label{fig:tsp_nine}
  \end{subfigure}
  \caption{Pathological example of \textit{lexicographic}'s suboptimality compared to TSP.}
  \label{fig:pathological}
\end{figure}

\textit{\textbf{Advantage Over Lexicographic Ordering --}}
Although the
\textit{lexicographic} ordering often leads to large amounts of gate cancellation, it does not achieve the optimality of TSP over the entire Hamiltonian. As an example, consider the 4-qubit Hamiltonian with lexicographically ordered strings $[XXXX$, $XXYY$, $XYXY$, $XYYX$, $YXXY$, $YXYX$, $YYXX$, $YYYY]$. These eight commuting strings arise in the Jordan-Wigner encoding for molecules, so this example is ubiquitous \cite{gokhale2019minimizing, whitfield2011simulation}. Applying Eq.~\eqref{eq:cnot_upper_bound}, we see that $8 \times 2 \times 4 = 64$ CNOTs are needed prior to gate cancellation. After CNOT cancellation, summing Eq.~\eqref{eq:cnot_distance} along the lexicographic order gives a total of 40 CNOTs. Now consider the TSP order $[XXXX$, $XXYY$, $XYXY$, $XYYX$, $YXYX$, $YXXY$, $YYXX$, $YYYY]$, which flips the fifth and sixth terms from the lexicographic order. Under TSP, we are able to generate circuits with only 36 CNOTs. See Fig.~\ref{fig:cancellation} for an example of the gate cancellations that are available in these two cases. In summary, unoptimized to \textit{lexicographic} to TSP have CNOT costs of $64 \to 40 \to 36$.

In certain cases, TSP can have an even greater factor of improvement over \textit{lexicographic} ordering. As an example, consider the nine Pauli terms in Fig.~\ref{fig:pathological}, sorted by \textit{lexicographic} and TSP orderings.

Without gate cancellation, $9 \times 10 \times 2 = 180$ CNOTs are required. Under the \textit{lexicographic} order, gate cancellation yields 112 CNOTs. However, with reordering into the TSP route, only 62 CNOTs are needed. While this particular example is pathological, it demonstrates scenarios where \textit{lexicographic} ordering is asymptotically identical to no-gate cancellation, but TSP achieves an asymptotic advantage. This advantage is from $O(N^2)$ to $O(N \log{N})$, which is a significant improvement.

\begin{figure}[t!]
    \centering
    \includegraphics[width=0.95\columnwidth]{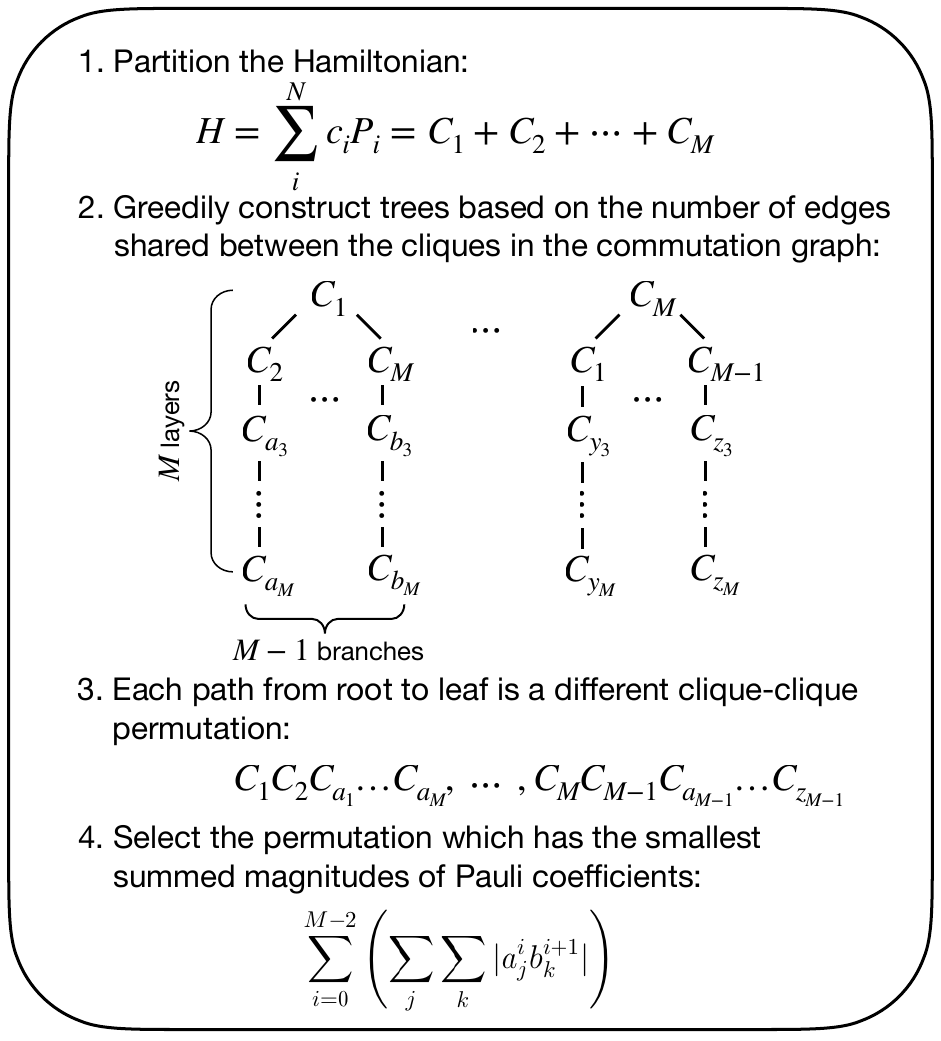}
    \caption{Polynomial time heuristic for selecting a clique-clique ordering. (1) The commutation graph, which was constructed to partition the Hamiltonian into $M$ cliques, is (2) utilized to greedily construct many trees based on the number of edges shared between cliques. (3) Each traversal of a tree from root to leaf produces a different clique-clique permutation. (4) The permutation with the smallest commutator magnitude (Eq.~\eqref{eqn:clique-commutator}) is selected.}
    \label{fig:permutation_heuristic}
\end{figure}

\subsection{Ordering the Cliques}\label{sec:permutation_heuristic}
Finally, the last step in \textit{max-commute-tsp} is to select the order in which the cliques, produced via the minimum clique cover from Sec.~\ref{sec:mincliquecover}, are simulated. Different orderings of the cliques can incur different Trotter errors by the same reasoning given above for the varying Pauli term orderings. Determining the optimal clique-clique ordering is intractable as it would require knowledge of the nested commutators between each of the cliques after the Baker–Campbell–Hausdorff expansion, i.e., an exponential amount of computation. Therefore, we use a heuristic, which runs in polynomial time, to decide upon a clique-clique ordering that relies on the approximated commutator: the first-order approximated difference between $U_{approx}$ and $U_{exact}$ after the Taylor expansion. Note that, for two cliques $C_1 = \sum^{N_1}_{i=1} a_{i} A_{i}$ and $C_2 = \sum^{N_2}_{i=1} b_{i} B_{i}$, where $a_i, b_i \in \mathbb{R}$ and $A_i, B_i$ are Pauli terms, the commutator between them is
\begin{equation}
    [C_1, C_2] = C_1 C_2 - C_2 C_1 = \sum_{i=1}^{N_1} \sum_{j=1}^{N_2} a_i b_j (A_i B_j - B_j A_i).
    \label{eqn:clique-commutator}
\end{equation}


Rather than compute the commutator between cliques exactly, the heuristic, shown in Fig.~\ref{fig:permutation_heuristic}, exploits the information stored in the commutation graph that was used to group the Pauli terms into fully commuting cliques. Counting the number of edges between $C_1$ and $C_2$ in the commutator graph indicates the number of terms within Eq.~\eqref{eqn:clique-commutator}'s sum that evaluate to zero. The heuristic uses the number of inter-clique edges to greedily grow a tree (the nodes of the tree representing cliques) where each path through the tree corresponds to a separate clique-clique ordering. Intuitively, a permutation selected in this manner will produce commutators between the consecutive cliques that contain many zero terms. One can hope that this will reduce the overall magnitude of the commutator and therefore contribute very little to the overall Trotter error. 

For each clique produced in the minimum clique cover, the heuristic constructs the tree described above with the current clique as the root. It then traverses each of these trees, producing a set of possible clique-clique permutations. Finally, for each permutation, compute $\sum_{i=1}^{N_1} \sum_{j=1}^{N_2} |a_i b_j|$ over the non-zero terms and select the permutation with the smallest value, which one would expect to have the smallest contribution to the Trotter error. For a Hamiltonian which is partitioned into $M$ cliques, this heuristic has a runtime of $O(M^4)$ and produces $O(M^2)$ different permutations.

\section{Methodology}\label{sec:methodology}
We compare the performance of different ordering strategies by measuring their ability to mitigate both physical and algorithmic errors via simulation and real device executions.
A benchmark set of 79 molecular Hamiltonians was generated using the OpenFermion software package \cite{mcclean2017openfermion} and the NIST Chemistry WebBook \cite{nist1997chemistry}.

For each of the ordering strategies we simulate the time dynamics of the benchmark Hamiltonians and at every time step we increase the Trotter number $r$ until the HS achieves an error $\epsilon < 0.1$.
The accuracy of a Hamiltonian simulation is measured using the diamond distance~\cite{aharonov1998quantum, gilchrist2005distance, campbell2019random}. The diamond distance between two quantum processes $\mathcal{E}$ and $\mathcal{F}$ is defined as:  
\begin{equation}
\begin{split}
    d_\Diamond(\mathcal{E}, \mathcal{F}) := \epsilon &= \lVert \mathcal{E} - \mathcal{F} \rVert_\Diamond \\
   &= \max_\rho \lVert (\mathcal{E} \otimes \mathbbm{1}) \rho - (\mathcal{F} \otimes \mathbbm{1}) \rho \rVert_1,
\end{split}
\label{eqn:diamond_distance}
\end{equation}
where $\rho$ is the density matrix representation of a quantum state, $\mathbbm{1}$ is the identity operator acting on the same size Hilbert space as $\mathcal{E}$ and $\mathcal{F}$, and $\lVert \dots \rVert_1$ denotes the trace norm. The diamond distance is an important and commonly used metric for distinguishing between two quantum processes in the absence of noise (i.e., considering only algorithmic errors)~\cite{campbell2019random, watrous2018theory}. Once the error threshold is met we report the number of CNOT gates required in the final quantum circuit. The results of these noiseless simulations are presented in Sec.~\ref{sec:noiseless_simulation}.

To capture the combined effects of algorithmic and physical errors on hardware execution we use noisy simulations with a depolarizing error model $\mathcal{E} (\rho) = \frac{pI}{2} + (1 - p)\rho$ which adds a noise channel on the two-qubit gates \cite{nielsen2002quantum}. Each entangling gate has probability $p$ of depolarizing its control and target qubits (replacing the qubits with the completely mixed state $I/2$), and probability $1 - p$ of leaving the qubits untouched.

We compare the output distributions of the noisy simulations and hardware executions using the Hellinger infidelity \cite{harper2020efficient, viacheslav2005operational, qiskit2020hellinger}. The Hellinger infidelity is defined as $1 - H_D(P,Q)$, where $H_D(P,Q)$ is the Hellinger distance, defined for two probability distributions $P$ and $Q$ as
\begin{equation}
    H_D (P,Q) = \frac{1}{\sqrt{2}} ||\sqrt{P}-\sqrt{Q}||_2.
    \label{eqn:hellinger}
\end{equation}

\begin{figure*}[t]
\centering
    \begin{subfigure}{0.535\textwidth}
    \includegraphics[width=\linewidth]{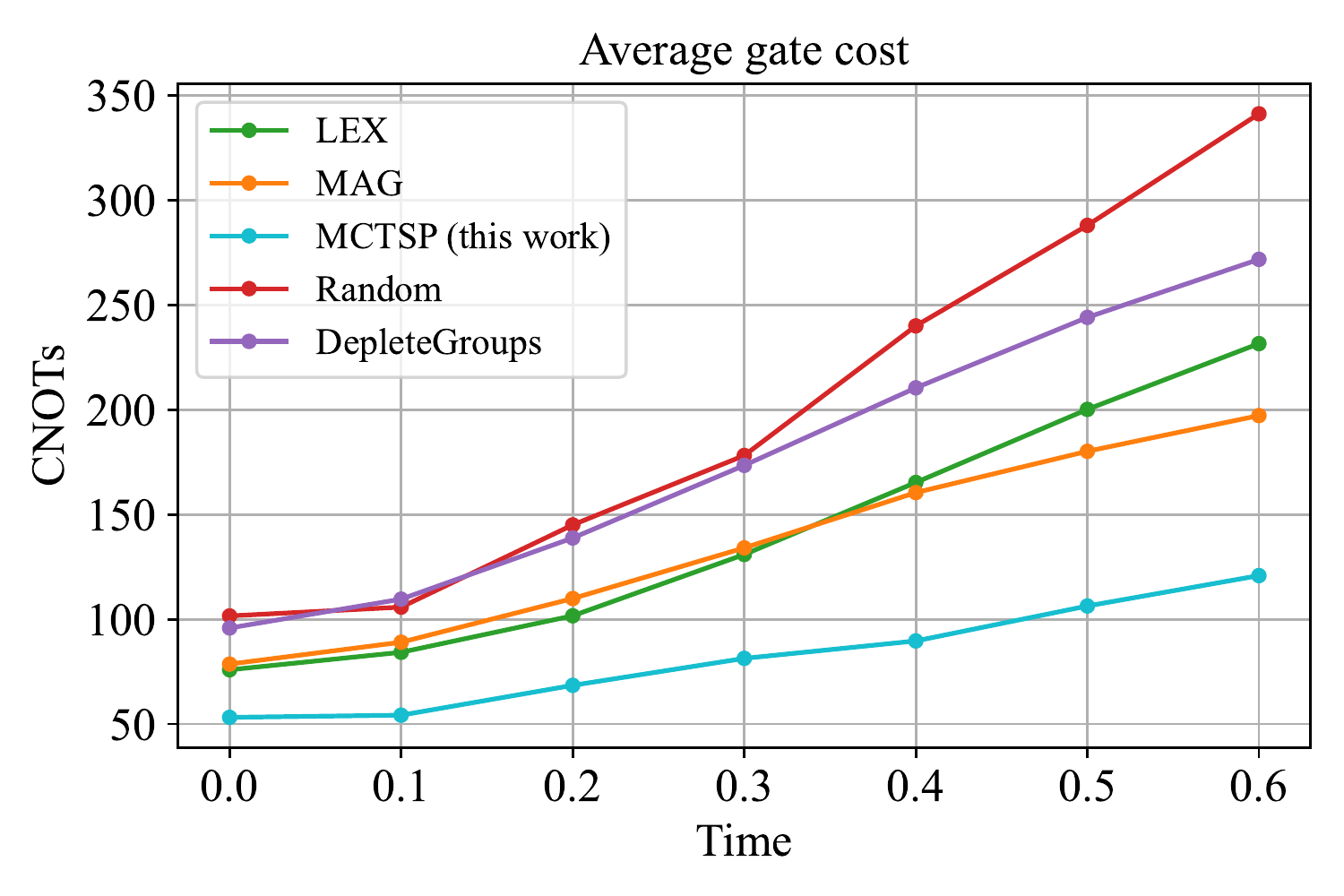}
    \vspace*{-5mm}
    \caption{\hspace*{-4em}}
    \label{fig:gatecost}
    \end{subfigure}
    \hfill
    \begin{subfigure}{0.45\textwidth}
    \includegraphics[width=\linewidth]{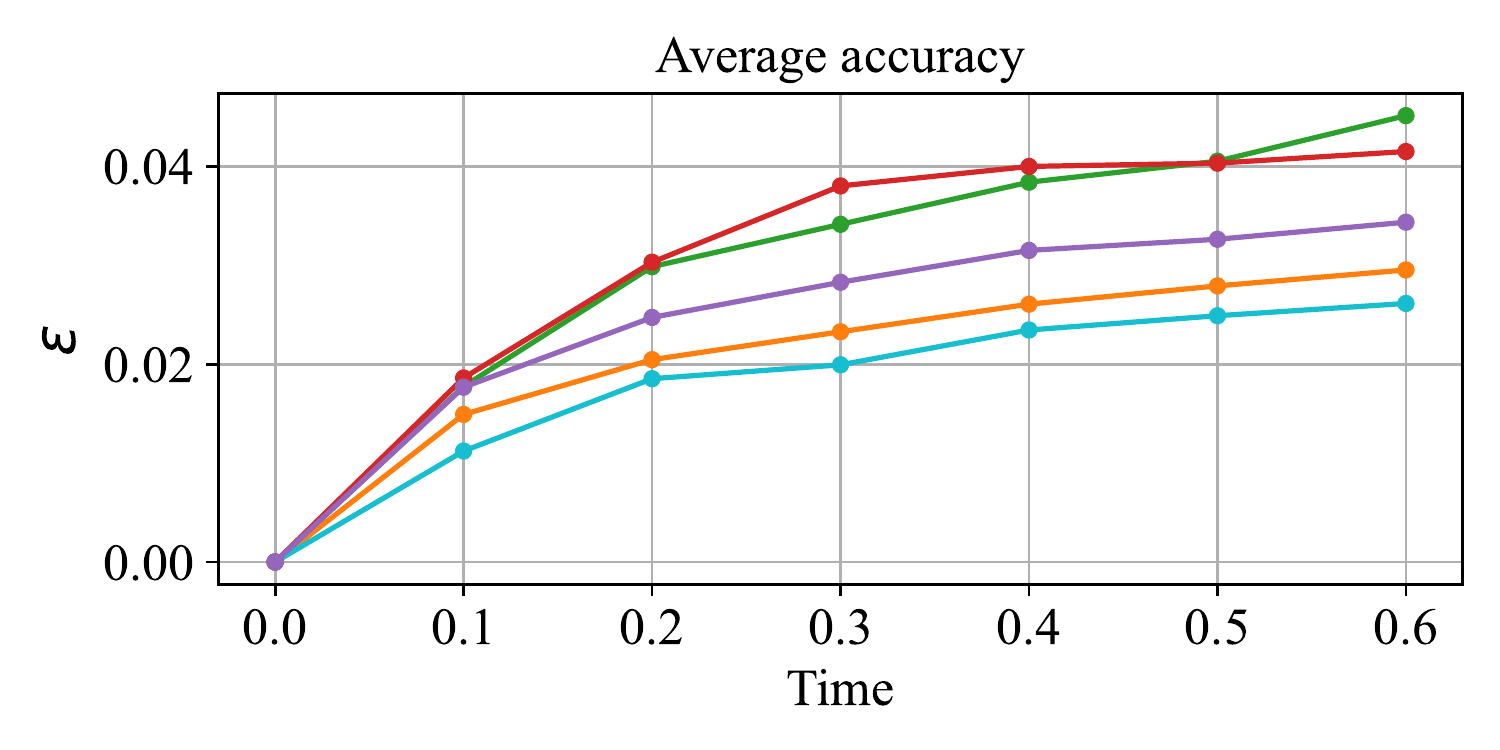}
    \vspace*{-5mm}
    \caption{\hspace*{-3.5em}}
    \label{fig:accuracy}
    \vspace*{+2mm}
    \includegraphics[width=\linewidth]{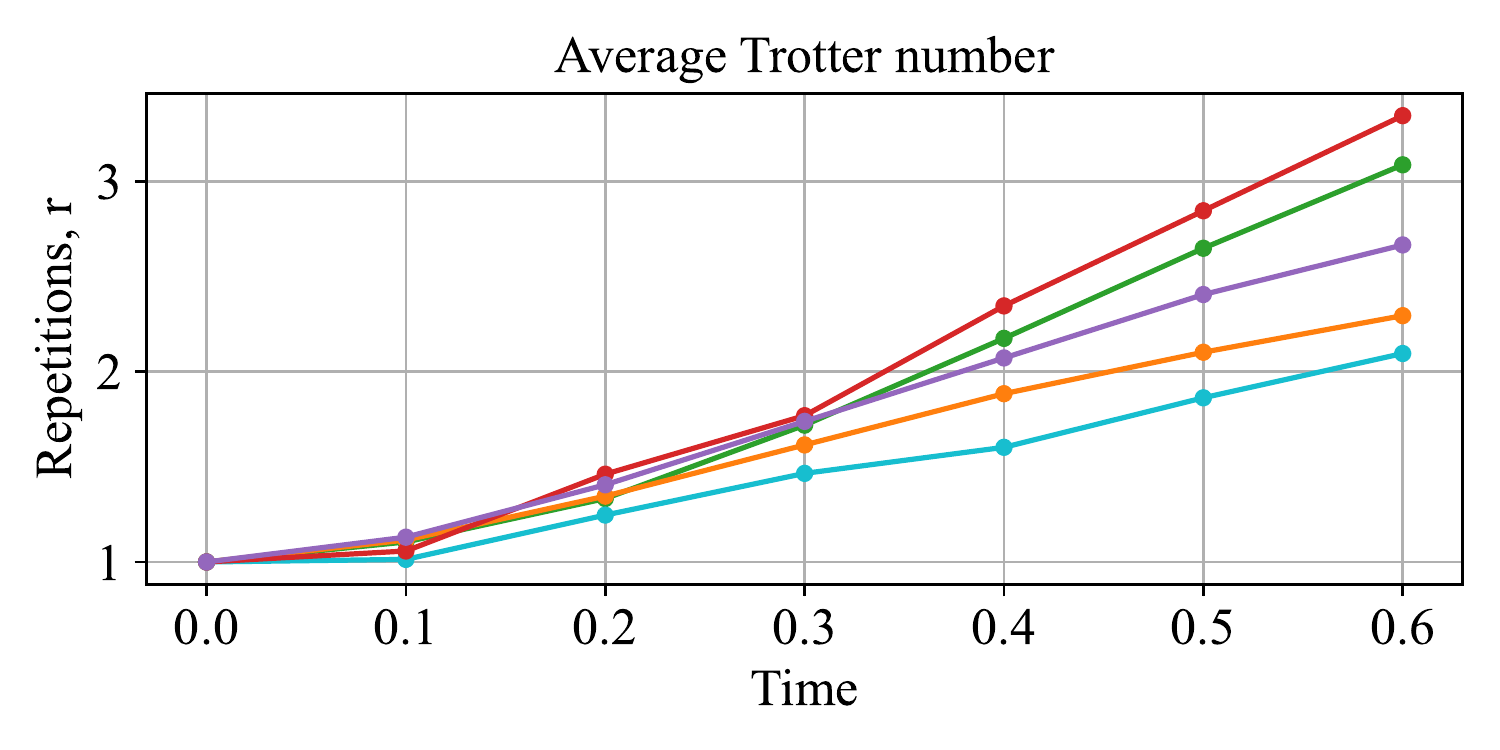}
    \vspace*{-5mm}
    \caption{\hspace*{-2.5em}}
    \label{fig:trotternumber}
    \end{subfigure}
\caption{Noiseless simulations of the time dynamics of 79 molecular Hamiltonians with an error threshold $\epsilon < 0.1$. Each molecule is represented with 4 qubits, and at every time step the Trotter number $r$ is increased until the diamond distance between the HS quantum circuit and the exact evolution unitary is below the error threshold.}
\label{fig:megaresults}
\end{figure*}
Eq.~\eqref{eqn:hellinger} is preferable for these experiments because they output probability distributions. Additionally, the diamond distance involves a maximization over all quantum states $\rho$ to capture the worst-case difference between two quantum channels, but for experiments on real hardware a specific initial state must be chosen. 

All of the code used to generate and test the benchmarks is available online in a Github repository~\cite{anonymous2020github}.



\section{Benchmark Results} \label{sec:benchmarks}

\subsection{Noiseless Simulation}\label{sec:noiseless_simulation}
In both the NISQ and fault-tolerant regimes, circuit depth and total gate count are important metrics of comparison for quantum circuits. In the NISQ era, these metrics are closely related to the probability that a circuit will execute successfully or succumb to the effects of noise. In the fault-tolerant regime, circuit depth and gate count are directly linked to the runtime of quantum programs. In the case of Hamiltonian simulation, a higher Trotter number $r$ provides a more accurate simulation, but the total gate cost scales proportionally with $r$.

In Fig.~\ref{fig:megaresults} we report the results of noiseless simulations of the time dynamics of the benchmark molecular Hamiltonians. At every time step we require that the HS achieve an error $\epsilon < 0.1$, increasing $r$ until this constraint is satisfied, and then computing the final number of CNOT gates required. In Fig.~\ref{fig:gatecost} \textit{max-commute-tsp} is able to simultaneously mitigate both algorithmic and physical errors. Compared to the other ordering strategies, \textit{max-commute-tsp} is able to produce accurate HS circuits using fewer entangling gates. On average, both the \textit{max-commute-tsp} and \textit{magnitude} orderings require similar Trotter numbers to reach the error threshold $\epsilon < 0.1$ as shown in Fig.~\ref{fig:trotternumber}. However, the TSP ordering of the Pauli terms within cliques allows the \textit{max-commute-tsp} strategy to cancel an average of 39\% more gates than the \textit{magnitude} ordering. The average 40\% difference in gate count between the \textit{lexicographic} and \textit{max-commute-tsp} strategies is mostly due to the higher Trotter numbers needed for the \textit{lexicographic} ordering to surpass the accuracy threshold.

The need to mitigate both sources of error is made apparent by comparing the results of the \textit{lexicographic} and \textit{magnitude} orderings. The \textit{lexicographic} HS circuits cancel many gates and are able to mitigate physical errors quite well. The \textit{magnitude} ordering is well suited to the regime of molecular Hamiltonians and is able to achieve high accuracy with low Trotter number. In Fig.~\ref{fig:gatecost}, at small $t$ values the poor accuracy of the \textit{lexicographic} ordering is compensated by its ability for gate cancellation, but for $t > 0.35$ the \textit{magnitude} ordering is able to produce shorter HS circuits because it requires a smaller Trotter number.

\subsection{Noisy Simulation and Hardware Evaluation}\label{sec:noisy_results}
We conducted noisy simulations of each benchmark Hamiltonian under a depolarizing error model using the \textit{lexicographic}, \textit{magnitude}, and \textit{max-commute-tsp} orderings with an initial state $\psi_{init} = \frac{1}{\sqrt{2}}(\ket{0011}+\ket{1100})$. In addition, six small benchmark Hamiltonians were evaluated on trapped ion processors. For each simulation and hardware execution the time and Trotter number parameters were set to $t = 1$ and $r = 1$ due to the error rates of current NISQ hardware. The increased noise incurred by the deeper circuits for $r > 1$ quickly overwhelms the results of the computation.

\begin{figure}[t]
    \centering
    \includegraphics[width=1.0\columnwidth]{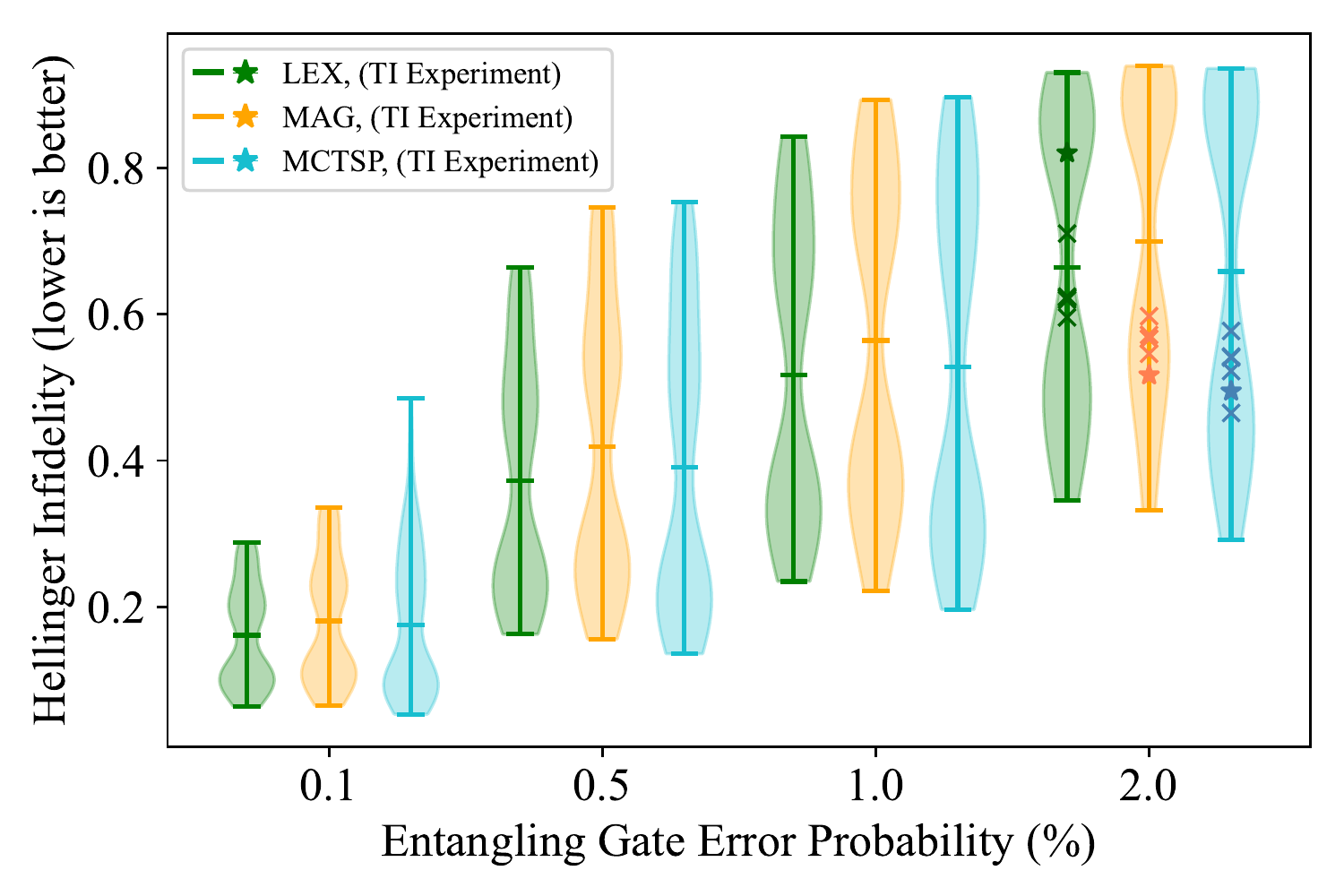}
    \caption{Distribution of Hellinger infidelities after evolving the entangled initial state $\psi_{init}$. The solid lines making up each violin plot denote the max, mean, and min of the distribution. The stars and crosses indicate experimental results obatained via a 7- and 11-qubit ion trap processor, respectively~\cite{alderete2020quantum, wright2019benchmarking}.}
    \label{fig:ghz_init}
\end{figure}

Fig.~\ref{fig:ghz_init} shows the measured distribution of Hellinger infidelities across four different error rates. The \textit{max-commute-tsp} ordering produces HS circuits with lower infidelity on average and also attains a minimum infidelity that is 1.2, 2.4, 3.6, 6.7\% lower than the minimum achieved by either the \textit{lexicographic} or \textit{magnitude} orderings for each of the 0.1, 0.5, 1.0, 2.0\% error rates, respectively. 

The results of the trapped ion experiments are contained in Table~\ref{tab:trappedion_experiment}. The ethene ($C_2H_4$) benchmark was run on a 7-qubit device~\cite{alderete2020quantum} (4000 shots) while the remainder utilized an IonQ device with 11-qubits~\cite{wright2019benchmarking} (1000 shots each). Each benchmark circuit was prepared in the initial state $\psi_{init}$ and consists of 4 data qubits and 1 ancilla qubit. Both processors have average entangling gate errors near 2\%. The experimental results are in shown in Fig.~\ref{fig:ghz_init} as the stars (7-qubit) and crosses (11-qubit). The hardware experiments confirm the results of the noisy-simulations and also highlight the impact of algorithmic errors --- even for current NISQ processors. For the $Cl_2$, $C_2H_2$, $F_2$, and $N_2$ benchmark molecules the \textit{magnitude} ordering, which produced deeper quantum circuits, attained lower infidelities than the \textit{lexicographic} ordering. The difference in performance can be attributed to the high (low) diamond distance between the \textit{lexicographic} (\textit{magnitude}) ordering and the ideal evolution. Interestingly, both strategies produced circuits of equal depth for the $O_2$ benchmark and achieved similar Hellinger infidelities despite a large difference in diamond distance. This may be due to the specific choice of initial state since the diamond distance is a measure of error over all initial states while the reported Hellinger infidelity is measured with respect to a single initial state.

\begin{table}[t]
\begin{tabular}{cccc}
\multicolumn{1}{l}{}                    & \multicolumn{1}{l}{}                   & \multicolumn{1}{l}{}               & \multicolumn{1}{l}{}                       \\ \hline
\multicolumn{1}{|c|}{Molecule}          & \multicolumn{1}{c|}{Diamond Dist. ($\epsilon$)} &
\multicolumn{1}{c|}{2-qubit gates} & \multicolumn{1}{c|}{Hellinger Inf. (\%)} \\
\multicolumn{1}{|c|}{}          & \multicolumn{1}{c|}{\textit{lex}, \textit{mag}, \textit{mctsp}} &
\multicolumn{1}{c|}{\textit{lex}, \textit{mag}, \textit{mctsp}} & \multicolumn{1}{c|}{\textit{lex}, \textit{mag}, \textit{mctsp}} \\
\hline
\multicolumn{1}{|c|}{\textit{$C_2H_4$}} & \multicolumn{1}{c|}{2.9, 1.8e-3, 1.8e-3}    & \multicolumn{1}{c|}{55, 49, 41}    & \multicolumn{1}{c|}{75.9, 55.2, \textbf{53.8}}      \\
\multicolumn{1}{|c|}{\textit{$Cl_2$}}   & \multicolumn{1}{c|}{1.9, 1.4e-4, 1.4e-4}    & \multicolumn{1}{c|}{47, 53, 37}    & \multicolumn{1}{c|}{62.2, 56.6, \textbf{54.2}}      \\
\multicolumn{1}{|c|}{\textit{$C_2H_2$}} & \multicolumn{1}{c|}{1.9, 1.3e-3, 1.3e-3}    & \multicolumn{1}{c|}{47, 53, 37}    & \multicolumn{1}{c|}{62.4, \textbf{57.2}, 57.7}      \\
\multicolumn{1}{|c|}{\textit{$F_2$}}    & \multicolumn{1}{c|}{1.1, 3.0e-3, 3.0e-3}    & \multicolumn{1}{c|}{47, 53, 37}    & \multicolumn{1}{c|}{71.0, 56.7, \textbf{52.2}}      \\
\multicolumn{1}{|c|}{\textit{$N_2$}}    & \multicolumn{1}{c|}{1.7, 1.2e-3, 1.2e-3}    & \multicolumn{1}{c|}{47, 53, 37}    & \multicolumn{1}{c|}{61.9, 54.6, \textbf{54.1}}      \\
\multicolumn{1}{|c|}{\textit{$O_2$}}    & \multicolumn{1}{c|}{3.1, 4.1e-3, 4.1e-3}    & \multicolumn{1}{c|}{41, 41, 27}    & \multicolumn{1}{c|}{59.6, 59.7, \textbf{46.5}}      \\ \hline
\end{tabular}
\caption{Diamond distances, gate counts, and Hellinger infidelities for the benchmarks evaluated on ion trap QPUs.} 
\label{tab:trappedion_experiment}
\end{table}

\section{Conclusion \& Future Directions}\label{sec:conclusion}
In this work, we introduced a new compilation method for Pauli term ordering, \textit{max-commute-tsp}, which is able to \textbf{simultaneously mitigate algorithmic and physical errors} in the quantum circuits that perform Hamiltonian simulations. Our hardware-level insights into the nature of the algorithmic errors guided us toward a solution which exploits the commutativity between Pauli terms to mitigate these errors. Additionally, \textit{max-commute-tsp} is able to mitigate physical errors by observing that the Pauli terms within each clique can be reordered to maximize gate cancellation via a TSP heuristic. While maintaining the same accuracy, \textit{max-commute-tsp} is able to produce HS circuits that are 39\% shorter than other compilation methods. Additionally, we use realistic noise models and real device executions to demonstrate the combined importance of algorithmic and physical error mitigation. 

It is important to note that, in the case of Hamiltonian simulation, even small improvements in accuracy can have significant impact. We point out two specific reasons below:
\begin{itemize}
    \item When concatenating many instances of the HS circuits together (i.e., for HS with Trotter number $r>1$), errors will accumulate with each repetition of the circuit. Therefore, reducing the error of the HS subcircuit can have an exponential impact on the final accuracy. For variational algorithms (such as VQE and QAOA), the same circuit is executed many times with classical post-processing in between. Reducing the error within this circuit will at least linearly increase the accuracy of any estimated observables.
    \item For quantum chemistry applications, a small reduction in error is likely to have a substantial effect. An improvement in quantum circuit fidelity would similarly improve a quantum algorithm's ability to compute exact ground state energies. Most classical approaches utilize approximation algorithms to estimate the ground state energy (since an exact diagonalization requires exponential classical computing resources) and even single digit improvements in accuracy are highly desirable~\cite{douglas2019electronic, mccaskey2019quantum}.
\end{itemize}

Our open-source software repository~\cite{anonymous2020github} allows users to construct HS circuits compiled via any of the ordering strategies considered here. This general compilation framework can be easily adapted to specific problem instances. For example, one can perform \textit{max-commute-tsp} on a partial set of the Pauli terms if the user has prior knowledge that the partial sets tend to share similar physical properties.  

Future work will involve developing Hamiltonian simulations for regimes other than the molecular systems studied here (e.g., solid-state structures\cite{marzari2012maximally}, high energy physics\cite{Bauer2019a}, protein folding\cite{robert2019resource}, etc.). The varied physical properties of these systems may result in different correlations between Pauli terms. These regimes may by difficult for inflexible strategies such as \textit{magnitude} or \textit{lexicographic} because most of their development has been centered around molecular Hamiltonians. Recent work on QAOA \cite{stuart2017from} suggests the use of complicated mixing Hamiltonians that consist of many non-commuting Pauli terms, which would invoke Trotter errors during evolution. A flexible ordering strategy like \textit{max-commute-tsp} would significantly reduce such Trotter errors, while also keeping the circuits short, improving the performance of QAOA and expanding the domain of tractable problems.

\section*{Acknowledgment}
The authors would like to thank Nathan Wiebe for the introduction to HS and for helpful discussions on commuting Pauli terms, Pavel Lougovski for initial thoughts and helpful discussions, and Yipeng Huang, Xiaoliang Wu and Zain Saleem for helpful comments. We also want to thank Alaina Green, Norbert Linke and the rest of the UMD group for access and help with running the 7-qubit ion trap experiments.

This work is funded in part by EPiQC, an NSF Expedition
in Computing, under grants CCF-1730082/1730449; in part
by STAQ under grant NSF Phy-1818914; in part by NSF
Grant No. 2110860; in part by the US Department of Energy Office 
of Advanced Scientific Computing Research, Accelerated 
Research for Quantum Computing Program; and in part by 
NSF OMA-2016136 and in part based upon work supported by the 
U.S. Department of Energy, Office of Science, National Quantum 
Information Science Research Centers. This work is also supported in part by DOE grants DE-SC0020289 and DE-SC0020331, and the Q-NEXT DOE NQI Center.

Y.S. was also funded in part by the NSF QISE-NET fellowship under grant number 1747426. The work of K.G. and M.S. is based upon work supported by the U.S. Department of Energy, Office of Science, Office of Fusion Energy Sciences, under Award Number DE-SC0020249. This work was also funded by the US Department of Energy Office, Advanced Manufacturing Office (CRADA No. 2020-20099.) and by the NSF under Grant No. 2110860.

\section*{Conflicts of Interest}
Fred Chong is Chief Scientist at Super.tech and an advisor to Quantum Circuits,~Inc.


\bibliographystyle{unsrt}
\bibliography{refs}

\begin{thebibliography}{10}

\bibitem{arute2019quantum}
F.~Arute, K.~Arya, R.~Babbush, D.~Bacon, Joseph~C. Bardin, R.~Barends,
  R.~Biswas, S.~Boixo, Fernando G. S.~L. Brand{\~a}o, David~A. Buell, Brian~J.
  Burkett, Y.~Chen, Z.~Chen, B.~Chiaro, R.~Collins, W.~Courtney, A.~Dunsworth,
  E.~Farhi, B.~Foxen, A.~Fowler, C.~Gidney, M.~Giustina, R.~Graff, Keith
  Guerin, Steve Habegger, Matthew~P. Harrigan, M.~Hartmann, A.~Ho, M.~Hoffmann,
  T.~Huang, T.~Humble, S.~V. Isakov, E.~Jeffrey, Z.~Jiang, D.~Kafri,
  K.~Kechedzhi, J.~Kelly, P.~Klimov, S.~Knysh, A.~Korotkov, F.~Kostritsa,
  D.~Landhuis, Mike Lindmark, E.~Lucero, D.~Lyakh, Salvatore Mandr{\`a},
  J.~McClean, M.~McEwen, A.~Megrant, X.~Mi, K.~Michielsen, M.~Mohseni,
  J.~Mutus, O.~Naaman, M.~Neeley, C.~Neill, M.~Y. Niu, E.~Ostby, A.~Petukhov,
  John~C. Platt, C.~Quintana, E.~Rieffel, P.~Roushan, N.~Rubin, D.~Sank,
  K.~Satzinger, V.~Smelyanskiy, Kevin~J. Sung, Matthew~D. Trevithick,
  A.~Vainsencher, Benjamin Villalonga, T.~White, Z.~Yao, P.~Yeh, Adam Zalcman,
  H.~Neven, and J.~Martinis.
\newblock Quantum supremacy using a programmable superconducting processor.
\newblock {\em Nature}, 574:505--510, 2019.

\bibitem{corcoles2019challenges}
Antonio~D C{\'o}rcoles, Abhinav Kandala, Ali Javadi-Abhari, Douglas~T McClure,
  Andrew~W Cross, Kristan Temme, Paul~D Nation, Matthias Steffen, and
  JM~Gambetta.
\newblock Challenges and opportunities of near-term quantum computing systems.
\newblock {\em arXiv preprint arXiv:1910.02894}, 2019.

\bibitem{wright2019benchmarking}
Kwame-Lante Wright, K.~Beck, S.~Debnath, J.~Amini, Y.~Nam, N.~Grzesiak,
  Jiah-Shing Chen, N.~C. Pisenti, M.~Chmielewski, C.~Collins, K.~Hudek,
  J.~Mizrahi, J.~D. Wong-Campos, S.~Allen, J.~Apisdorf, P.~Solomon,
  M.~Williams, A.~M. Ducore, A.~Blinov, S.~Kreikemeier, V.~Chaplin, M.~Keesan,
  C.~Monroe, and J.~Kim.
\newblock Benchmarking an 11-qubit quantum computer.
\newblock {\em Nature Communications}, 10, 2019.

\bibitem{barends2016digitized}
R.~Barends, A.~Shabani, L.~Lamata, J.~Kelly, A.~Mezzacapo, U.~Las Heras,
  R.~Babbush, A.~Fowler, B.~Campbell, Y.~Chen, Z.~Chen, B.~Chiaro,
  A.~Dunsworth, E.~Jeffrey, E.~Lucero, A.~Megrant, J.~Mutus, M.~Neeley,
  C.~Neill, P.~O’Malley, C.~Quintana, P.~Roushan, D.~Sank, A.~Vainsencher,
  J.~Wenner, T.~White, E.~Solano, H.~Neven, and J.~Martinis.
\newblock Digitized adiabatic quantum computing with a superconducting circuit.
\newblock {\em Nature}, 534 7606:222--6, 2016.

\bibitem{hempel2018quantum}
Cornelius Hempel, C.~Maier, Jonathan Romero, J.~McClean, T.~Monz, H.~Shen,
  P.~Jurcevic, B.~Lanyon, P.~Love, R.~Babbush, Al{\'a}n Aspuru-Guzik, R.~Blatt,
  and C.~Roos.
\newblock Quantum chemistry calculations on a trapped-ion quantum simulator.
\newblock {\em Physical Review X}, 8, 2018.

\bibitem{peruzzo2014variational}
Alberto Peruzzo, Jarrod McClean, Peter Shadbolt, Man-Hong Yung, Xiao-Qi Zhou,
  Peter~J Love, Al{\'a}n Aspuru-Guzik, and Jeremy~L O’brien.
\newblock A variational eigenvalue solver on a photonic quantum processor.
\newblock {\em Nature communications}, 5:4213, 2014.

\bibitem{tomesh2020coreset}
Teague Tomesh, Pranav Gokhale, Eric~R Anschuetz, and Frederic~T Chong.
\newblock Coreset clustering on small quantum computers.
\newblock {\em Electronics}, 10(14):1690, 2021.

\bibitem{lloyd1996universal}
Seth Lloyd.
\newblock Universal quantum simulators.
\newblock {\em Science}, pages 1073--1078, 1996.

\bibitem{feynman1982simulating}
Richard~P Feynman.
\newblock Simulating physics with computers.
\newblock {\em Int. J. Theor. Phys}, 21(6/7), 1982.

\bibitem{childs2018toward}
Andrew~M Childs, Dmitri Maslov, Yunseong Nam, Neil~J Ross, and Yuan Su.
\newblock Toward the first quantum simulation with quantum speedup.
\newblock {\em Proceedings of the National Academy of Sciences},
  115(38):9456--9461, 2018.

\bibitem{mueck2015quantum}
Leonie Mueck.
\newblock Quantum reform.
\newblock {\em Nature chemistry}, 7(5):361--363, 2015.

\bibitem{aspuru2005simulated}
Al{\'a}n Aspuru-Guzik, Anthony~D Dutoi, Peter~J Love, and Martin Head-Gordon.
\newblock Simulated quantum computation of molecular energies.
\newblock {\em Science}, 309(5741):1704--1707, 2005.

\bibitem{lidar1999calculating}
Daniel~A Lidar and Haobin Wang.
\newblock Calculating the thermal rate constant with exponential speedup on a
  quantum computer.
\newblock {\em Physical Review E}, 59(2):2429, 1999.

\bibitem{reiher2017elucidating}
Markus Reiher, Nathan Wiebe, Krysta~M Svore, Dave Wecker, and Matthias Troyer.
\newblock Elucidating reaction mechanisms on quantum computers.
\newblock {\em Proceedings of the National Academy of Sciences},
  114(29):7555--7560, 2017.

\bibitem{preskill2018quantum}
John Preskill.
\newblock Quantum computing in the nisq era and beyond.
\newblock {\em Quantum}, 2:79, 2018.

\bibitem{appel2004modern}
Andrew~W Appel.
\newblock {\em Modern compiler implementation in C}.
\newblock Cambridge university press, 2004.

\bibitem{babbush2015chemical}
Ryan Babbush, Jarrod McClean, Dave Wecker, Al{\'a}n Aspuru-Guzik, and Nathan
  Wiebe.
\newblock Chemical basis of trotter-suzuki errors in quantum chemistry
  simulation.
\newblock {\em Physical Review A}, 91(2):022311, 2015.

\bibitem{childs2019faster}
Andrew~M Childs, Aaron Ostrander, and Yuan Su.
\newblock Faster quantum simulation by randomization.
\newblock {\em Quantum}, 3:182, 2019.

\bibitem{tran2020destructive}
Minh~C Tran, Su-Kuan Chu, Yuan Su, Andrew~M Childs, and Alexey~V Gorshkov.
\newblock Destructive error interference in product-formula lattice simulation.
\newblock {\em Physical Review Letters}, 124(22):220502, 2020.

\bibitem{tranter2019ordering}
Andrew Tranter, Peter~J Love, Florian Mintert, Nathan Wiebe, and Peter~V
  Coveney.
\newblock Ordering of trotterization: Impact on errors in quantum simulation of
  electronic structure.
\newblock {\em Entropy}, 21(12):1218, 2019.

\bibitem{tranter2018comparison}
Andrew Tranter, Peter~J Love, Florian Mintert, and Peter~V Coveney.
\newblock A comparison of the bravyi--kitaev and jordan--wigner transformations
  for the quantum simulation of quantum chemistry.
\newblock {\em Journal of chemical theory and computation}, 14(11):5617--5630,
  2018.

\bibitem{hastings2014improving}
Matthew~B Hastings, Dave Wecker, Bela Bauer, and Matthias Troyer.
\newblock Improving quantum algorithms for quantum chemistry.
\newblock {\em arXiv preprint arXiv:1403.1539}, 2014.

\bibitem{suzuki2005}
Naomichi Hatano and Masuo Suzuki.
\newblock Finding exponential product formulas of higher orders.
\newblock In Arnab Das and Bikas Chakrabarti, editors, {\em Quantum Annealing
  and Related Optimization Methods}, pages 37--67. Springer, Berlin, 2005.

\bibitem{mcclean2016theory}
Jarrod~R McClean, Jonathan Romero, Ryan Babbush, and Al{\'a}n Aspuru-Guzik.
\newblock The theory of variational hybrid quantum-classical algorithms.
\newblock {\em New Journal of Physics}, 18(2):023023, 2016.

\bibitem{van2020circuit}
Ewout van~den Berg and Kristan Temme.
\newblock Circuit optimization of hamiltonian simulation by simultaneous
  diagonalization of pauli clusters.
\newblock {\em Quantum}, 4:322, 2020.

\bibitem{gokhale2019minimizing}
Pranav Gokhale, Olivia Angiuli, Yongshan Ding, Kaiwen Gui, Teague Tomesh,
  Martin Suchara, Margaret Martonosi, and Frederic~T Chong.
\newblock {$ O (N^3) $ Measurement Cost for Variational Quantum Eigensolver on
  Molecular Hamiltonians}.
\newblock {\em IEEE Transactions on Quantum Engineering}, 1:1--24, 2020.

\bibitem{verteletskyi2020measurement}
Vladyslav Verteletskyi, Tzu-Ching Yen, and Artur~F Izmaylov.
\newblock Measurement optimization in the variational quantum eigensolver using
  a minimum clique cover.
\newblock {\em The Journal of Chemical Physics}, 152(12):124114, 2020.

\bibitem{yen2020measuring}
Tzu-Ching Yen, Vladyslav Verteletskyi, and Artur~F Izmaylov.
\newblock Measuring all compatible operators in one series of single-qubit
  measurements using unitary transformations.
\newblock {\em Journal of Chemical Theory and Computation}, 16(4):2400--2409,
  2020.

\bibitem{nist1997chemistry}
Peter Linstrom.
\newblock {NIST Chemistry WebBook, NIST Standard Reference Database 69}, 1997.

\bibitem{mcclean2017openfermion}
Jarrod~R McClean, Nicholas~C Rubin, Kevin~J Sung, Ian~D Kivlichan, Xavier
  Bonet-Monroig, Yudong Cao, Chengyu Dai, E~Schuyler Fried, Craig Gidney,
  Brendan Gimby, et~al.
\newblock Openfermion: the electronic structure package for quantum computers.
\newblock {\em Quantum Science and Technology}, 5(3):034014, 2020.

\bibitem{incite}
US~Department of~Energy.
\newblock {DOE INCITE} program, 2020.
\newblock https://www.doeleadershipcomputing.org/awardees/.

\bibitem{jordan1928pauli}
Pascual Jordan and Eugene~P Wigner.
\newblock About the pauli exclusion principle.
\newblock {\em Z. Phys.}, 47:631--651, 1928.

\bibitem{bravyi2002fermionic}
Sergey~B Bravyi and Alexei~Yu Kitaev.
\newblock Fermionic quantum computation.
\newblock {\em Annals of Physics}, 298(1):210--226, 2002.

\bibitem{childs2019theory}
Andrew~M Childs, Yuan Su, Minh~C Tran, Nathan Wiebe, and Shuchen Zhu.
\newblock Theory of trotter error with commutator scaling.
\newblock {\em Physical Review X}, 11(1):011020, 2021.

\bibitem{poulin2014trotter}
David Poulin, Matthew~B Hastings, Dave Wecker, Nathan Wiebe, Andrew~C Doherty,
  and Matthias Troyer.
\newblock The trotter step size required for accurate quantum simulation of
  quantum chemistry.
\newblock {\em arXiv preprint arXiv:1406.4920}, 2014.

\bibitem{raeisi2012quantum}
Sadegh Raeisi, Nathan Wiebe, and Barry~C Sanders.
\newblock Quantum-circuit design for efficient simulations of many-body quantum
  dynamics.
\newblock {\em New Journal of Physics}, 14(10):103017, 2012.

\bibitem{whitfield2011simulation}
James~D Whitfield, Jacob Biamonte, and Al{\'a}n Aspuru-Guzik.
\newblock Simulation of electronic structure hamiltonians using quantum
  computers.
\newblock {\em Molecular Physics}, 109(5):735--750, 2011.

\bibitem{nielsen2002quantum}
Michael~A. Nielsen and Isaac~L. Chuang.
\newblock {\em Quantum Computation and Quantum Information: 10th Anniversary
  Edition}.
\newblock Cambridge University Press, USA, 10th edition, 2011.

\bibitem{alam2020circuit}
Mahabubul Alam, Abdullah Ash-Saki, and Swaroop Ghosh.
\newblock Circuit compilation methodologies for quantum approximate
  optimization algorithm.
\newblock In {\em 2020 53rd Annual IEEE/ACM International Symposium on
  Microarchitecture (MICRO)}, pages 215--228. IEEE, 2020.

\bibitem{itoko2019quantum}
Toshinari Itoko, Rudy Raymond, Takashi Imamichi, Atsushi Matsuo, and Andrew~W
  Cross.
\newblock Quantum circuit compilers using gate commutation rules.
\newblock In {\em Proceedings of the 24th Asia and South Pacific Design
  Automation Conference}, pages 191--196, 2019.

\bibitem{knapp2005basic}
Anthony~W Knapp.
\newblock {\em Basic real analysis}.
\newblock Springer Science \& Business Media, 2005.

\bibitem{dollard1979product}
John~Day Dollard and Charles~N. Friedman.
\newblock {\em Product Integration with Application to Differential Equations}.
\newblock Cambridge University Press, USA, 1st edition, 1979.

\bibitem{karp1972reducibility}
Richard~M Karp.
\newblock Reducibility among combinatorial problems.
\newblock In {\em Complexity of computer computations}, pages 85--103.
  Springer, 1972.

\bibitem{boppana1992approximating}
Ravi Boppana and Magn{\'u}s~M Halld{\'o}rsson.
\newblock Approximating maximum independent sets by excluding subgraphs.
\newblock {\em BIT Numerical Mathematics}, 32(2):180--196, 1992.

\bibitem{leighton1979graph}
Frank~Thomson Leighton.
\newblock A graph coloring algorithm for large scheduling problems.
\newblock {\em Journal of research of the national bureau of standards},
  84(6):489--506, 1979.

\bibitem{planat2007pauli}
Michel Planat and Metod Saniga.
\newblock On the pauli graphs of n-qudits.
\newblock {\em arXiv preprint quant-ph/0701211}, 2007.

\bibitem{debnath2016demonstration}
Shantanu Debnath, Norbert~M Linke, Caroline Figgatt, Kevin~A Landsman, Kevin
  Wright, and Christopher Monroe.
\newblock Demonstration of a small programmable quantum computer with atomic
  qubits.
\newblock {\em Nature}, 536(7614):63, 2016.

\bibitem{linke2017experimental}
Norbert~M Linke, Dmitri Maslov, Martin Roetteler, Shantanu Debnath, Caroline
  Figgatt, Kevin~A Landsman, Kenneth Wright, and Christopher Monroe.
\newblock Experimental comparison of two quantum computing architectures.
\newblock {\em Proceedings of the National Academy of Sciences},
  114(13):3305--3310, 2017.

\bibitem{sahni1976p}
Sartaj Sahni and Teofilo Gonzalez.
\newblock P-complete approximation problems.
\newblock {\em Journal of the ACM (JACM)}, 23(3):555--565, 1976.

\bibitem{christofides1976worst}
Nicos Christofides.
\newblock Worst-case analysis of a new heuristic for the travelling salesman
  problem.
\newblock Technical report, Carnegie-Mellon Univ Pittsburgh PA Management
  Sciences Research Group, 1976.

\bibitem{genova2017experimental}
Kyle Genova and David~P Williamson.
\newblock An experimental evaluation of the best-of-many {Christofides’}
  algorithm for the traveling salesman problem.
\newblock {\em Algorithmica}, 78(4):1109--1130, 2017.

\bibitem{aharonov1998quantum}
Dorit Aharonov, Alexei Kitaev, and Noam Nisan.
\newblock Quantum circuits with mixed states.
\newblock In {\em Proceedings of the thirtieth annual ACM symposium on Theory
  of computing}, pages 20--30, 1998.

\bibitem{gilchrist2005distance}
Alexei Gilchrist, Nathan~K Langford, and Michael~A Nielsen.
\newblock Distance measures to compare real and ideal quantum processes.
\newblock {\em Physical Review A}, 71(6):062310, 2005.

\bibitem{campbell2019random}
Earl Campbell.
\newblock Random compiler for fast hamiltonian simulation.
\newblock {\em Physical review letters}, 123(7):070503, 2019.

\bibitem{watrous2018theory}
John Watrous.
\newblock {\em The theory of quantum information}, chapter 3.3.
\newblock Cambridge university press, 2018.

\bibitem{harper2020efficient}
Robin Harper, Steven~T Flammia, and Joel~J Wallman.
\newblock Efficient learning of quantum noise.
\newblock {\em Nature Physics}, pages 1--5, 2020.

\bibitem{viacheslav2005operational}
Viacheslav~P. Belavkin, Giacomo~Mauro D'Ariano, and Maxim Raginsky.
\newblock Operational distance and fidelity for quantum channels.
\newblock {\em Journal of Mathematical Physics}, 46(6):062106, 2005.

\bibitem{qiskit2020hellinger}
Qiskit hellinger fidelity, 2020.
\newblock Available at
  \url{https://qiskit.org/documentation/stubs/qiskit.quantum_info.hellinger_fidelity.html#qiskit.quantum_info.hellinger_fidelity}.

\bibitem{anonymous2020github}
Teague Tomesh.
\newblock {dqs-term-grouping}.
\newblock \url{https://github.com/teaguetomesh/dqs-term-grouping}, 2021.

\bibitem{alderete2020quantum}
C~Huerta Alderete, Shivani Singh, Nhung~H Nguyen, Daiwei Zhu, Radhakrishnan
  Balu, Christopher Monroe, CM~Chandrashekar, and Norbert~M Linke.
\newblock Quantum walks and dirac cellular automata on a programmable
  trapped-ion quantum computer.
\newblock {\em Nature Communications}, 11(1):1--7, 2020.

\bibitem{douglas2019electronic}
Oscar~A Douglas-Gallardo, David~Adrian Saez, Stefan Vogt-Geisse, and Esteban
  V{\"o}hringer-Martinez.
\newblock Electronic structure benchmark calculations of inorganic and
  biochemical carboxylation reactions.
\newblock {\em Journal of computational chemistry}, 40(13):1401--1413, 2019.

\bibitem{mccaskey2019quantum}
Alexander~J McCaskey, Zachary~P Parks, Jacek Jakowski, Shirley~V Moore, Titus~D
  Morris, Travis~S Humble, and Raphael~C Pooser.
\newblock Quantum chemistry as a benchmark for near-term quantum computers.
\newblock {\em npj Quantum Information}, 5(1):1--8, 2019.

\bibitem{marzari2012maximally}
Nicola Marzari, Arash~A. Mostofi, Jonathan~R. Yates, Ivo Souza, and David
  Vanderbilt.
\newblock Maximally localized wannier functions: Theory and applications.
\newblock {\em Rev. Mod. Phys.}, 84:1419--1475, Oct 2012.

\bibitem{Bauer2019a}
Benjamin Nachman, Davide Provasoli, Wibe~A de~Jong, and Christian~W Bauer.
\newblock Quantum algorithm for high energy physics simulations.
\newblock {\em Physical Review Letters}, 126(6):062001, 2021.

\bibitem{robert2019resource}
Anton Robert, Panagiotis~Kl Barkoutsos, Stefan Woerner, and Ivano Tavernelli.
\newblock Resource-efficient quantum algorithm for protein folding.
\newblock {\em npj Quantum Information}, 7(1):1--5, 2021.

\bibitem{stuart2017from}
Stuart Hadfield, Zhihui Wang, Bryan O’Gorman, Eleanor~G Rieffel, Davide
  Venturelli, and Rupak Biswas.
\newblock From the quantum approximate optimization algorithm to a quantum
  alternating operator ansatz.
\newblock {\em Algorithms}, 12(2):34, 2019.

\end{thebibliography}

\end{document}